\documentclass{LMCS}

\usepackage[british]{babel}
\usepackage[T1]{fontenc}
\usepackage[ansinew]{inputenc}
\usepackage[all]{xypic}
\usepackage{amsmath,amssymb}
\usepackage{Common/prooftree}
\usepackage{hyperref,enumerate}
\usepackage{epic} 
\usepackage{eepic} 
\usepackage{graphicx}

\long\def\ignore#1{\relax}

\newcommand\struto{\raise6pt\hbox{\strut}}%
\newcommand\strutb{\raise-6pt\hbox{\strut}}%

\newcommand\upline{\hline\struto}
\newcommand\midline{\\[+4pt]\hline\struto}
\newcommand\mmidline{\\[+4pt]\hline\hline\struto}
\newcommand\downline{\\[+4pt]\hline}

\theoremstyle{definition}
\theoremstyle{definition}\newtheorem{example}[thm]{Example}
\theoremstyle{definition}

\theoremstyle{plain}
\theoremstyle{plain}

\renewcommand{\paragraph}[1]{\mbox{}\\{\bf \large #1}}

\newcommand\infers[3]{\infer[#3]{#2}{#1}}
\newcommand\Infers[3]{\begin{prooftree}#1\Justifies #2\using #3\end{prooftree}}

\newcommand\dom[1]{\textsf{Dom}{(#1)}}

\newcommand{\es}{\emptyset}
\newcommand{\eqdef}{:=\ }

\def\FV#1{\textsf{FV}(#1)}
\newcommand{\sep}{\mbox{$\;|\;$}}    

\renewcommand\l{\lambda}

\newcommand{\Gam}{\Gamma}

\newcommand{\ie}{i.e.\ }
\newcommand{\eg}{e.g.\ }
\newcommand{\cf}{cf.\ }

\newcommand{\resp}{resp.\ }

\def\cut{\textsf{cut}}
\newcommand\NJ{\textsf{NJ}}

\newcommand\Giii{\textsf{G3}}

\def\lb{$\overline{\lambda}$}

\def\CoC{{\textsf{CoC}}}

\def\name{{\textsf{SCoC}^U}}
\def\nameg{{\textsf{PTSC}}}
\def\PTS{{\textsf{PTS}}}
\def\PTSC{{\textsf{PTSC}}}
\def\PTSCa{{\PTSC\ensuremath{\alpha}}}
\def\PTSa{{\PTS\ensuremath{\alpha}}}
\def\LJT{{\textsf{LJT}}}
\def\Coq{{\textsf{Coq}}}
\def\Lego{{\textsf{Lego}}}

\def\Ltac{\ensuremath{\mathcal{L}_\textit{tac}}}
\def\Lpdt{\ensuremath{\mathcal{L}_\textit{pdt}}}

\newcommand\CRS{{\textsf{CRS}}}
\newcommand\ERS{{\textsf{ERS}}}

\newcommand{\subst}[3]{ \left\{{}^{#3}\hspace{-6pt}\diagup\hspace{-2pt}_{#2} \right\}\hspace{-1pt} #1 }

\newcommand{\Rew}[1]{\longrightarrow_{#1}\;}
\newcommand{\Rewn}[1]{{\longrightarrow^{*}}_{#1}\;}
\newcommand{\Rewplus}[1]{{\longrightarrow^{+}}_{#1}\;}
\newcommand{\Rewsn}[1]{{\longleftrightarrow^{*}}_{#1}\;}

\def\SN#1{\textsf{SN}^{#1}}

\newcommand\blob{\star}
\newcommand\un[1]{\textsf{i}(#1)}
\newcommand\deux[2]{\textsf{ii}(#1,#2)}

\newcommand\focut[3][{}]{{\textsf{cut}^{#1}(#2,#3)}}
\newcommand\fosub[3][{}]{{\textsf{sub}^{#1}(#2,#3)}}
\newcommand\fotuple[2][n]{{\textsf{tuple}^{#1}(#2)}}

\newcommand\unname{\textsf{i}}
\newcommand\deuxname{\textsf{ii}}

\newcommand\focutname[1][{}]{{\textsf{cut}^{#1}}}
\newcommand\fosubname[1][{}]{{\textsf{sub}^{#1}}}
\newcommand\fotuplename[1][n]{{\textsf{tuple}^{#1}}}

\def\foe#1{{\mathcal S}(#1)}

\newcommand\fot[1]{{\mathcal T}(#1)}

\newcommand\pre{\prec}
\newcommand\glpo{>_{\textsf{lpo}}}

\def\SVar{{\mathcal X}}
\def\SCons{{\mathcal S}}
\def\STerms{{\mathcal T}}
\def\SLists{{\mathcal L}}

\def\K{{\Box}}
\def\T{{\star}}
\def\Rel{{\mathcal R}}
\def\Sort{{\mathcal A}}

\def\P#1#2#3{\Pi {#1}^{#2}.{#3}}
\def\lami#1#2#3{\lambda {#1}^{#2}.{#3}}
\def\cont#1#2{{#1}\;{#2}}

\def\el{[]}
\def\conc#1#2{{#1}@{#2}}
\def\st#1#2{{#1}\!\cdot\!{#2}}

\def\cuti#1#2#3#4{\langle{#2}/{#3}\rangle{#4}}

\def\B{\textsf{B}}
\def\Ai{\textsf{A1}}
\def\Aii{\textsf{A2}}
\def\Aiii{\textsf{A3}}
\def\Aiv{\textsf{A4}}
\def\Bi{\textsf{B1}}
\def\Bii{\textsf{B2}}
\def\Biii{\textsf{B3}}

\def\Ci{\textsf{C1}}
\def\Cii{\textsf{C2}}
\def\Ciii{\textsf{C3}}
\def\Civ{\textsf{C4}}
\def\Cv{\textsf{C5}}
\def\Cvi{\textsf{C6}}
\def\Calpha{\textsf{C}\alpha}
\def\Di{\textsf{D1}}
\def\Dii{\textsf{D2}}
\def\Diii{\textsf{D3}}
\def\Dbeta{\textsf{D}\beta}

\def\As{\textsf{As}}
\def\Bs{\textsf{Bs}}
\def\Cs{\textsf{Cs}}

\def\Ds{\textsf{Ds}}

\def\xsubst{\textsf{xsubst}}
\def\x{\textsf{x}}

\def\Herb{\textsf{}}

\def\All{\textsf{}}

\def\redT{\Rew{\All}}

\ignore{

}

\newcommand\col{\!:\!}
\def\tpd{\!:\!}

\newcommand\arr{\hspace{-3pt}\rightarrow\hspace{-3pt}}

\newcommand{\seqf}[2][]{\mbox{$\ {\pmb \vdash}_{#1}^{#2}\ $}}

\newcommand{\seq}[1][]{\seqf[#1]{}}

\newcommand\Seqf[4][]{#3\seqf[{#1}]{#2} #4}
\newcommand\Seq[3][]{#2\seq[{#1}] #3}

\newcommand\Derif[5][]{\Seqf[#1]{#2}{#3}{{#4}\col{#5}}}
\newcommand\Derip[4]{\Seq[#1]{#2}{{#3}\col{#4}}}
\newcommand\Deri[4][]{\Derip{#1}{#2}{#3}{#4}}
\newcommand\Deris[5][]{\Derip{#1}{#2;#3}{#4}{#5}}
\newcommand\Derist[3]{\Derif\star{#1} {#2} {#3}}
\newcommand\Derisst[4]{\Derif\star{#1;{#2}} {#3} {#4}}

\def\wf{\hspace{5 pt}\textsf{wf}}

\newcommand\DeriPTSC[3]{\Derip{\PTSC\!}{#1}{#2}{#3}}
\newcommand\DeriPTSCs[4]{\Derip{\PTSC\!}{{#1};{#2}}{#3}{#4}}

\def\conv{\Rewsn{}}

\def\emptyenv{\emptyset}

\def\empt{\textsf{empty}}
\def\extend{\textsf{extend}}
\def\Ttyped{\textsf{sorted}}
\def\contr#1{\textsf{Select}_{#1}}
\def\ax{\textsf{axiom}}
\def\Picons{\Pi\textsf{wf}}
\def\Pir{\Pi\textsf{R}}
\def\Pil{\Pi\textsf{L}}
\def\convr{\textsf{conv}_R}
\def\convl{\textsf{conv}_L}
\def\cut{\textsf{Cut}}
\def\subenv{\sqsubseteq}

\newcommand\delambda[1][]{{\mathcal A}_{#1}}
\newcommand\alambda{{\mathcal B}}

\newcommand\kCCC[1]{{\delambda}{(#1)}}
\newcommand\kCC[2][]{{\delambda}_{#1}{(#2)}}
\newcommand\kCCname{{\delambda}}

\renewcommand\k[1]{\alambda{(#1)}}
\newcommand\kl[2]{\alambda^{#1}{(#2)}}
\newcommand\kname{{\mathcal B}}

\def\ke#1{\k{#1}}
\def\kename{\kname}
\def\kel#1#2{\kl{#1}{#2}}

\def\DPTS{\vdash_{\PTS}}
\def\DeriCoC#1#2#3{{#1}\DPTS {#2}\tpd{#3}}

\newcommand{\namePS}{\textsf{PS}}
\newcommand\DPS{\vdash_{\namePS}}
\newcommand\PSS{\namePS}

\newcommand\DeriPS[3]{{#1}\DPS {#2}\col{#3}}
\newcommand\DeriPSs[4]{{#1};{#2}\DPS {#3}\col{#4}}

\newcommand\PilPS{\ensuremath{\Pil_\namePS}}

\newcommand{\namePE}{\textsf{PE}}
\newcommand\enum[5]{\Deri{#2}{#4}{#3} {{{\pmb\sep} {#5}}}}
\newcommand\enums[6]{\Deris{#2}{#3}{#5}{#4}{{{\pmb\sep} {#6}}}}
\newcommand\goals[2]{#1\Longrightarrow #2}

\newcommand\PE{\namePE}

\newcommand\enumPS[5]{\Derip{\PE}{#2}{#4}{#3}\sep {#5}}
\newcommand\enumPSs[6]{\Derip{\PE}{#2;#3}{#5}{#4}\sep {#6}}
\newcommand\goalsPS[2]{#1\Longrightarrow_{\PE} #2}

\newcommand\constr[3]{#2\stackrel{#1}= #3}

\newcommand\TtypedPE{\textsf{sorted}_\namePE}
\newcommand\axPE{\textsf{ax}_\namePE}
\newcommand\PiconsPE{\Pi\textsf{wf}_\namePE}
\newcommand\PirPE{\ensuremath{\Pir_\namePE}}
\newcommand\PilPE{\ensuremath{\Pil_\namePE}}

\def\doi{7 (1:6) 2011}
\lmcsheading%
{\doi}
{1--35}
{}
{}
{Oct.\phantom03, 2009}
{Mar.~22, 2011}
{}

\begin{document}

\title[Proof Search in PTS]{A Focused Sequent Calculus Framework for\\ Proof-Search in Pure Type Systems} 



\author[S.~Lengrand]{St{\'e}phane Lengrand\rsuper a}
\address{{\lsuper a}CNRS, {\'E}cole Polytechnique, France}
\email{Lengrand@LIX.Polytechnique.fr}

\author[R.~Dyckhoff]{Roy Dyckhoff\rsuper b}
\address{{\lsuper b}School of Computer Science, University of St Andrews, Scotland}
\email{rd@cs.st-andrews.ac.uk}

\author[J.~McKinna]{James McKinna\rsuper c}
\address{{\lsuper c}Radboud University, Nijmegen, The Netherlands}
\email{james.mckinna@cs.ru.nl}


\begin{abstract}

  Basic proof-search tactics in logic and type theory can be seen as
  the root-first applications of rules in an appropriate sequent
  calculus, preferably without the redundancies generated by
  permutation of rules. 
  This paper addresses the issues of defining such sequent calculi for
  \emph{Pure Type Systems} (\PTS, which were originally presented in
  natural deduction style) and then organizing their rules for effective
  proof-search.
  We introduce the idea of \emph{Pure Type Sequent Calculus with
    meta-variables} (\PTSCa), by enriching the syntax of a
  permutation-free sequent calculus for propositional logic due to
  Herbelin, which is strongly related to natural deduction and already
  well adapted to proof-search. The operational semantics is adapted
  from Herbelin's and is defined by a system of local rewrite rules as
  in cut-elimination, using explicit substitutions.  We prove
  confluence for this system.
  Restricting our attention to \PTSC, a type system for the
  \emph{ground} terms of this system, we obtain the \emph{Subject
    Reduction} property and show that each \PTSC\ is logically
  equivalent to its corresponding \PTS, and the former is strongly
  normalising iff the latter is.
  We show how to make the logical rules of \PTSC\ into a
  syntax-directed system \namePS\ for proof-search, by incorporating the
  conversion rules as in syntax-directed presentations of the \PTS\
  rules for type-checking.
  Finally, we consider how to use the explicitly scoped meta-variables
  of \PTSCa\ to represent partial proof-terms, and use them to analyse
  interactive proof construction.  This sets up a framework \namePE\
  in which we are able to study proof-search strategies, type
  inhabitant enumeration and (higher-order) unification.
  

\end{abstract}


\maketitle

\newpage
\tableofcontents

\section*{Introduction}

{\em Pure Type Systems (\PTS)} (see \eg\cite{Bar:intgts}) were independently introduced by Berardi~\cite{Berardi1988} and Terlouw~\cite{Terlouw1989} as a generalisation of Barendregt's $\lambda$-cube, and form a convenient framework for representing a range of different extensions of the simply-typed $\lambda$-calculus. \emph{System $F$}, \emph{System $F_\omega$}~\cite{Girard72}, \emph{System $\lambda\Pi$}~\cite{VanDaalen80,harper87framework}, and the \emph{Calculus of Constructions (\CoC)}~\cite{CoC} are examples of such systems, on which several major proof assistants are based (\eg \Coq~\cite{Coq}, \Lego~\cite{lego}, and the \emph{Edinburgh Logical Framework}~\cite{harper87framework}; \emph{Higher-Order Logic} can also be presented as a \PTS, but this is \emph{not} the basis of its principal implementation~\cite{HOL}).

With typed $\lambda$-calculus as their basis, such systems are traditionally presented in natural deduction style, with rules introducing and eliminating logical constants (aka type constructors). Dowek~\cite{Doweksynthesis} and Mu\~noz~\cite{Munozsynthesis} show how to perform proof-search in this style, 
   by enumerating type inhabitants. 

This however misses out on the advantages of sequent calculus~\cite{Gentzen35} for proof-search. As suggested by Plotkin~\cite{Plotkin87}, a Gentzen-style sequent calculus (with left and right introduction rules) can be used as a basis for proof-search in the case of $\lambda \Pi$~\cite{PymWallen91,PymSLogica95} (later extended to any \PTS~\cite{gutierrez02cutfree,gutierrez03elimination}).  However, the permutations of inference steps available in a Gentzen-style calculus (such as \Giii~\cite{Kle:intm}) introduce some extra non-determinism in proof-search.

Herbelin~\cite{Herbelin94,HerbThes} introduced a {\em permutation-free} calculus \LJT\ for intuitionistic logic, exploiting the focusing ideas of Andreoli \cite{andreoli92focusing}, Danos \textit{et al.} \cite{DJS95} and (ultimately) ideas from Girard's linear logic \cite{girard-ll}. Herbelin's calculus has been considered as a basis for proof-search in intuitionistic logic~\cite{dyckhoff99permutability}, generalising the uniform proof approach to logic programming (see~\cite{Miller91apal} for hereditary Harrop logic). A version with cut rules and proof-terms forms an explicit substitution calculus \lb~\cite{Herbelin94,DyckUrbJLC} with a strong connection to (call-by-name) $\beta$-reduction and abstract machines such as that of Krivine~\cite{KrivineMachine}.

This builds, as in the Curry-Howard correspondence, a computational interpretation of sequent calculus proofs on the basis of which type theory can be reformulated, 
   now 
with a view to formalising proof-search.  
   In earlier work~\cite{LDMcK06,LengrandPhD}, we reformulated the language and proof theory of \PTS s in terms of {\em Pure Type Sequent Calculi (\PTSC)}. The present paper completes this programme, introducing {\em Pure Type Sequent Calculi with meta-variables (\PTSCa)}, together with an operationalisation of proof-search in \PTS\ in terms of \PTSCa. 
It follows earlier work~\cite{pinto00sequent}, relating \lb\ to proof-search in the $\Lambda \Pi$ calculus~\cite{PymWallen91,PymSLogica95}.
   Introducing 
meta-variables for proof-search is the main technical novelty of this paper over~\cite{LDMcK06}.

This gives a secure but simple theoretical basis for the implementation of \PTS-based systems such as \Coq~\cite{Coq} and \Lego~\cite{lego}; these proof assistants feature interactive proof construction methods using proof-search tactics. As observed by~\cite{McKinnaLEGO}, the primitive tactics are 
   not in exact correspondence 
with the elimination rules of the underlying natural deduction formalism: while the tactic \textsf{intro} {\em does} correspond to the right-introduction rule for $\Pi$-types (whether in natural deduction or in sequent calculus), the tactics \textsf{apply} in \Coq{} and \textsf{Refine} in \Lego{}, however, are much closer (in spirit) to the left-introduction rule \(\Pil\) for $\Pi$-types in the focused sequent calculus \LJT\  than to the $\Pi$-elimination rule in natural deduction. 
   The \(\Pil\) rule types the construct $\st M l$ of \lb, representing a list of terms with head $M$ and tail $l$: 
   $$\infers {\Deri \Gamma M A\quad\Deris {\Gamma} {\cuti A M x B} {l} C} 
   {\Deris \Gamma {\P x A {B}} {\st M l} C} {\Pil} $$

However, the aforementioned tactics are also able to postpone the investigation of the first premiss and start investigating the second. This leads to incomplete proof-terms and unification constraints to be solved. Here, we integrate these features into \PTSC\ using explicitly scoped meta-variables. 
The resulting framework, called \PTSCa, supports the analysis and definition of interactive proof construction tactics (as in \Coq{} and \Lego{}), as well as type inhabitant enumeration (see \cite{Doweksynthesis,Munozsynthesis}).

Of course, formalising proof-search mechanisms has already been investigated, if only to design tactic languages like Delahaye's \Ltac\ and \Lpdt~\cite{delahaye}. Also noteworthy here are McBride's and Jojgov's PhD theses~\cite{mcbride,jojgov}, which consider extensions of type theory to admit partial proof objects. Using meta-variables similar to ours, Jojgov shows how to manage explicitly their progressive instantiation via a definitional mechanism and compares this with Delahaye's \Ltac\ and \Lpdt.

While formalising the connections with this line of research remains as future work, the novelty of our approach here is to use the sequent calculus to bridge the usual gap (particularly wide for \PTS\ and their implementations) between the rules defining a logic and the rules describing proof-search steps.  A by-product of this bridge is ensuring correctness of proof-search, whose output thus need not be type-checked (which it currently is, in most proof assistants).

One reason why this is possible in our framework is that it can decompose (and thus account for) some mechanisms that are usually externalised and whose outputs usually need to be type-checked, such as unification (including higher-order~\cite{HuetThEtat}). Indeed, it integrates the idea, first expounded in~\cite{Doweksynthesis}, that proof-search and unification generalise in type theory to a single process.

The rules of our framework may not be deterministic enough to be considered as specifying an algorithm, but they are atomic enough to provide an operational semantics in which algorithms such as the above can be specified. They thus provide a semantics not only for type inhabitation algorithms, but also more generally for tactic languages, and, more originally, for unification algorithms.

As an example, 
we consider commutativity of conjunction expressed in (the \PTSCa\ corresponding to) System $F$, previously presented in~\cite{LDMcK06} without meta-variables. We show here how meta-variables improve the formalisation of proof-search.

\ignore{As the trust in software eventually rely on the confidence in the code
matching the inference rules
with the hope to improve the understanding
of implementation of proof systems based on \PTS{} in  respect of:
\begin{itemize}
\item
a direct analysis of the basic tactics, which could then be moved into
the kernel, rather than requiring a separate type-checking layer for
correctness,
\item
opening the way to improve \ignore{abstract machines for reduction as
part of the basic system, for both reduction} the basic system with an
approach closer to abstract machines to express reductions, both in
type-checking \emph{and} in execution (of extracted programs),
\item
studying extensions to systems involving inductive types/families (such
as the Calculus of Inductive Constructions).
\end{itemize}

For instance, we believe that they could improve
semi-automated proof-search, allow simpler and neater presentations of such
expressive systems. 
}


   Our work may be compared with 
that of our predecessors as follows: \vskip 6pt
\noindent
\begin{tabular}{|c|c@{\ }c@{\ }c@{\ }c|}
\hline
&Type Theory&Inference rules&Proof-terms&\begin{tabular}c
Formalisation of\\incomplete proofs\\(by \eg meta-variables)\end{tabular}\\\hline
\cite{PymSLogica95}&$\lambda\Pi$&\Giii&$\l$&YES \\
\cite{Doweksynthesis}&$\CoC$&\NJ&$\l$&YES\\
\cite{pinto00sequent}&$\lambda\Pi(\Sigma)$ &\LJT&\lb &NO\\
\cite{gutierrez03elimination}&\PTS& \Giii &$\l$ & NO\\
\cite{jojgov}&$\lambda$\textsf{HOL} & \NJ &$\l$ & YES\\
This paper&   \PTS&\LJT&\lb&YES\\
\hline
\end{tabular}
\vskip 6pt

Note that, in contrast to~\cite{PymSLogica95,gutierrez02cutfree,gutierrez03elimination}, we use a focused sequent calculus (\LJT) instead of an unfocused one (\Giii). The former forces proof-search to be `goal-directed' in the tradition of logic programming and uniform proofs, while the latter is more relaxed and would accommodate saturation-based reasoning. Our choice here is motivated by a tighter connection with natural deduction and by the tactics currently used in proof assistants such as \Coq\ and \Lego. While~\cite{PymSLogica95} does identify permutations of inference rules which would allow the recovery of a goal-directed strategy, \cite{gutierrez03elimination} focuses instead on the elimination of a cut-rule which then sheds a surprising light on the open problem of Expansion Postponement~\cite{Gutierrez03bis}. 

Our move from \Giii\ to \LJT\ is also particularly convenient to capture the process of higher-order unification as a proof-search mechanism.
Pym and Wallen address proof-search~\cite{PymWallen91} in the
particular case of \(\lambda\Pi\), the type theory of the Edinburgh
Logical Framework, using a black-box higher-order unification
algorithm adapted from that of Huet. They discuss how well-typedness
of meta-variable instantiations computed by unification can be
exploited to control the search space.
Meanwhile no meta-variables (or similar technology
supporting unification) feature
in~\cite{gutierrez02cutfree,gutierrez03elimination}.

In any case, this line of research keeps a traditional $\lambda$-calculus syntax for proof-terms, which thus does not reflect the structure of proof trees.
We sought instead a formalism whose terms reflect how proofs and unifiers are constructed, and so moved from $\lambda$-calculus to \lb.

\ignore{
In any case, this line of research keeps 
   a traditional $\lambda$-calculus 
syntax for proof-terms, which thus does not reflect the structure of proof trees. 
   We sought instead a formalism whose terms reflect how proofs are constructed, and so moved to \lb. 


   Finally, the focused sequent calculus is much better suited to the
   process of higher-order unification. This was treated by Pym
   in a $\lambda$-calculus setting~\cite{DavidPym:Unif},
   following Huet, but not
   in~\cite{gutierrez02cutfree,gutierrez03elimination}, where
   meta-variables do not feature.

}

The paper's structure is as follows: 
Section~\ref{sec:synsem} presents
the syntax of \PTSCa{}, 
   the full language of terms and lists containing meta-variables,  
and gives the rewrite rules for normalisation.
Section~\ref{sec:confluence} relates this syntax with that of $\lambda$-calculus
  in \PTS\ style 
and thereby derives the confluence of the \PTSCa-calculus.
Section~\ref{sec:type} presents a parametric typing system \PTSC\ for
ground terms (\ie the restriction to \PTSCa-terms containing \emph{no} meta-variables), 
and states and proves properties such as {\em Subject Reduction}.
Section~\ref{sec:pts} establishes the correspondence between a
\nameg{} and the \PTS\ with the same parameters; we show type
preservation and the strong normalisation result.
Section~\ref{sec:proofsearch} discusses proof-search in a \PTSC{}.
Section~\ref{sec:enum} introduces the inference system for \PTSCa,
as a way to formalise incomplete proofs and operationalise proof-search.
Section~\ref{sec:example} shows the aforementioned example.
These are followed by a conclusion and discussion of directions for further work.

Some ideas and results of this paper (namely
Sections~\ref{sec:confluence},~\ref{sec:type} and~\ref{sec:pts}, which were 
already presented in~\cite{LDMcK06}) have been formalised and machine-checked in
the \Coq\ system~\cite{SilesPTSC} 
 using a de Bruijn index representation, as
   in 
\eg \cite{LengrandPhD}.


\section{Syntax and operational semantics of \PTSCa}
\label{sec:synsem}

\subsection{Syntax}
\label{subsec:syn}

We consider an extension (with type annotations) of the proof-term
syntax \lb\ of Herbelin's focused sequent calculus
\LJT~\cite{HerbThes}. As in \lb, 
   the grammar of \PTSCa\ features 
two syntactic categories: that of \emph{terms} and that of \emph{lists}.

The syntax 
depends on a given set $\SCons$ of \emph{sorts},
written $s,s',\ldots$, a denumerable set $\SVar$ of \emph{variables}, written
$x,y,z,\ldots$, and two denumerable sets of \emph{meta-variables}:
those for terms, written $\alpha,\alpha',\ldots$, and those for
lists, written $\beta,\beta',\ldots$. These meta-variables come with
an intrinsic notion of \emph{arity}.

\begin{defi}[Terms and Lists]
The set $\STerms_\Herb$ of \emph{terms} (denoted $M,N,P$,\ldots, $A, B,
\ldots$) and the set $\SLists_\Herb$ of \emph{lists} (denoted
$l,l',\ldots$) are inductively defined by: 

$$\begin{array}{rl} 
M,N,P,A,B &::=\P x A B\sep \;\lami x
A M\sep \;s\sep \;\cont x l\sep \;\cont M l\sep \;\cuti A M x N\sep\;\alpha(M_1,\ldots,M_n)\\ 
l,l'&::=\el\sep \;\st M l\sep \;\conc l {l'}\sep \;\cuti M M x l\sep\; \beta(M_1,\ldots,M_n)
\end{array}
$$
where $n$ is the arity of $\alpha$ and $\beta$.

The constructs $\P x A M$, $\lami x A M$, and $\cuti A N x M$ bind $x$ in $M$, and
$\cuti M M x l$ binds $x$ in $l$, thus defining the free variables of
a term $M$ (\resp a list $l$), denoted $\FV M$ (\resp $\FV l$), as
well as $\alpha$-conversion, issues of which are treated in the usual
way.  Note that
$\FV{\alpha(M_1,\ldots,M_n)}=\FV{\beta(M_1,\ldots,M_n)}=\bigcup_{i=1}^n\FV{M_n}$; 
see the discussion on meta-variables below. 
A term $M$ is {\em closed} if $\FV M=\emptyset$. 
As usual, let $A\arr B$ denote $\P x A B$ when $x\not\in \FV B$. 



Terms and lists without meta-variables are called \emph{ground terms}
and \emph{ground lists}, respectively. (Previously, these were just
called terms and lists in~\cite{LDMcK06}).
\end{defi}

Lists are used to represent sequences of arguments of a function; the
term $\cont x l$ (resp. $\cont M l$) represents the application of $x$
(resp. $M$) to the list of arguments $l$.  Note that a variable alone
is not a term; it must be applied to a list, possibly the empty list,
denoted $\el$.  The list $\st M l$ has head $M$ and tail $l$, with a
typing rule corresponding to the left-introduction of $\Pi$-types (\cf
Section~\ref{sec:type}). The following figure shows the generic
structure of a $\l$-term 
   $\lami {x_1} {}{\ldots\lami{x_p}{} {V\;M_1\ldots M_n}}$,  
and its \lb-representation 
   as the term 
   $\lami {x_1} {}{\ldots\lami{x_p}{} {\cont V (\st{M_1}{\ldots \st {M_n}\el})}}$, 
   as follows: 
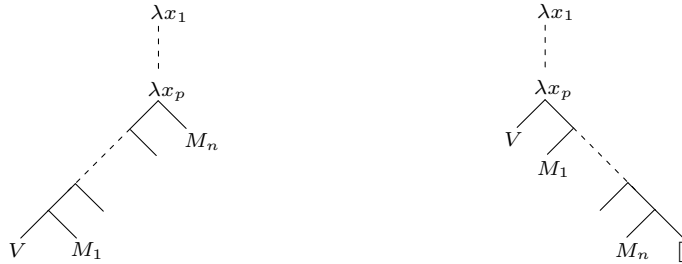
\begin{center}
\scalebox{1.1}{
\hfill  %
%
\setlength{\unitlength}{0.00050000in}
\begingroup\makeatletter\ifx\SetFigFont\undefined%
\gdef\SetFigFont#1#2#3#4#5{%
  \reset@font\fontsize{#1}{#2pt}%
  \fontfamily{#3}\fontseries{#4}\fontshape{#5}%
  \selectfont}%
\fi\endgroup%
{\renewcommand{\dashlinestretch}{30}
\begin{picture}(1846,2373)(0,-10)
%
%
\drawline(136,214)(398,476)(661,214)
%
%
\dashline{60.000}(585,664)(1172,1253)
%
%
\drawline(1173,1252)(1435,1514)(1698,1252)
%
%
\drawline(390,468)(652,730)(915,468)
%
%
\dashline{60.000}(1440,2218)(1440,1818)
%
%
\drawline(1173,1239)(1417,995)
\put(15,35){\makebox(0,0)[lb]{\smash{{\SetFigFont{7}{8.4}{\rmdefault}{\mddefault}{\updefault}$V$}}}}
\put(606,43){\makebox(0,0)[lb]{\smash{{\SetFigFont{7}{8.4}{\rmdefault}{\mddefault}{\updefault}$M_1$}}}}
\put(1680,1074){\makebox(0,0)[lb]{\smash{{\SetFigFont{7}{8.4}{\rmdefault}{\mddefault}{\updefault}$M_n$}}}}
\put(1360,2289){\makebox(0,0)[lb]{\smash{{\SetFigFont{7}{8.4}{\rmdefault}{\mddefault}{\updefault}$\lambda x_1$}}}}
\put(1335,1564){\makebox(0,0)[lb]{\smash{{\SetFigFont{7}{8.4}{\rmdefault}{\mddefault}{\updefault}$\lambda x_p$}}}}
\end{picture}
}


\hspace{3cm}
  %
%
\setlength{\unitlength}{0.00050000in}
\begingroup\makeatletter\ifx\SetFigFont\undefined%
\gdef\SetFigFont#1#2#3#4#5{%
  \reset@font\fontsize{#1}{#2pt}%
  \fontfamily{#3}\fontseries{#4}\fontshape{#5}%
  \selectfont}%
\fi\endgroup%
{\renewcommand{\dashlinestretch}{30}
\begin{picture}(1774,2386)(0,-10)
%
%
\drawline(1702,223)(1440,485)(1177,223)
%
%
\dashline{60.000}(1253,673)(666,1262)
%
%
\drawline(665,1261)(403,1523)(140,1261)
%
%
\drawline(1448,477)(1186,739)(923,477)
%
%
\dashline{60.000}(403,2231)(403,1831)
%
%
\drawline(669,1252)(425,1008)
\put(323,2302){\makebox(0,0)[lb]{\smash{{\SetFigFont{7}{8.4}{\rmdefault}{\mddefault}{\updefault}$\lambda x_1$}}}}
\put(298,1577){\makebox(0,0)[lb]{\smash{{\SetFigFont{7}{8.4}{\rmdefault}{\mddefault}{\updefault}$\lambda x_p$}}}}
\put(15,1089){\makebox(0,0)[lb]{\smash{{\SetFigFont{7}{8.4}{\rmdefault}{\mddefault}{\updefault}$V$}}}}
\put(323,814){\makebox(0,0)[lb]{\smash{{\SetFigFont{7}{8.4}{\rmdefault}{\mddefault}{\updefault}$M_1$}}}}
\put(1644,35){\makebox(0,0)[lb]{\smash{{\SetFigFont{7}{8.4}{\rmdefault}{\mddefault}{\updefault}$\el$}}}}
\put(1061,39){\makebox(0,0)[lb]{\smash{{\SetFigFont{7}{8.4}{\rmdefault}{\mddefault}{\updefault}$M_n$}}}}
\end{picture}
}


\hfill
  }
\end{center}

Successive applications give rise to list concatenation, denoted
$\conc l {l'}$ (with \(@\) acting as an explicit constructor). For
instance, the list $\conc {(\st{M_1}{\ldots \st {M_n}\el})}
{(\st{M_{n+1}}{\ldots \st {M_p}\el})}$ will reduce to $\st{M_1}{\ldots
  \st {M_n}{\st{M_{n+1}}{\ldots \st {M_p}\el}}}$.

The terms $\cuti K M x N$ and $\cuti K M x l$ are
explicit substitutions, on terms and lists, respectively. 
They will be used in two ways: first, to instantiate a universally
quantified variable, and second, to describe explicitly the
interaction between the constructors in the normalisation process
(given in Section~\ref{ssec:opsem}).

More intuition about Herbelin's calculus, its syntax and operational
semantics, may be found in~\cite{HerbThes}.

Among the features added to the syntax of \lb, our meta-variables can be seen as \emph{higher-order variables}.  
   As in \CRS~\cite{Klo80}, 
unknown terms are represented with (meta/higher-order) variables \emph{applied} to the series of (term-)variables that could occur freely in those terms, \eg $\alpha(x,y)$ (more formally, $\alpha(\cont x \el,\cont y \el)$) represents an unknown term $M$ in which $x$ and $y$ could occur free (and no other).  
Such arguments $x, y$ can later be 
   instantiated, 
so that $\alpha(N,P)$ represents $\subst M {x,y}{N,P}$.  
In other words, a meta-variable by itself stands for something closed, \ie a term under a series of bindings covering all its free variables, \eg  $x.y.M$ when $\FV M\subseteq \{x,y\}$ (using a traditional notation for higher-order terms, see \eg \cite{Terese03}, Ch.\ 11).\footnote{
   We develop this in Section~\ref{sec:enum} below. 
There is no binding mechanism for meta-variables in the syntax
of \PTSCa, but at the meta-level there is a natural notion of instantiation,
also presented in Section~\ref{sec:enum}. We thus emphasise the fact
that instantiation of meta-variables never occurs during computation;
in that respect, meta-variables really behave like constants
or term constructors.} 
This allows us to consider a simple notion of $\alpha$-conversion, with $\lami x s {\alpha(\cont x \el,\cont y \el)} \equiv_\alpha\lami z s {\alpha(\cont z \el,\cont y \el)}$. 
Henceforth, however, we will elide further discussion of such matters, and simply write \(=\) to denote \(\equiv_\alpha\).

This kind of meta-variable differs from that in~\cite{Munozsynthesis},
which is rather in the style of \ERS~\cite{Kha90} where the variables
that could occur freely in the unknown term are not specified
explicitly. The drawback of our approach is that we have to
\emph{know} in advance the free variables that might occur free in the
unknown term, but in a typed setting such as proof-search these are
actually the variables declared in the typing environment. Moreover,
although specifying explicitly the variables that could occur free in
an unknown term might seem heavy, it actually avoids the usual
(non-)confluence problems when terms contain meta-variables in the
style of~\ERS.\footnote{See the discussion at the end of
Section~\ref{sec:confluence}.} The solution in~\cite{Munozsynthesis}
has the drawback of not simulating $\beta$-reduction (although the
reductions reach the expected normal forms).

\subsection{Operational semantics}
\label{ssec:opsem}

The operational semantics of \PTSCa\ is given by the system of reduction rules 
in Figure~\ref{fig:red-rules}, 
comprising sub-systems $\B$, $\x'$, and $\xsubst'$, and combinations thereof. 
This system extends that of~\cite{LDMcK06} with rules $\Aiv,\Calpha,\Dbeta$. 
Side-conditions to avoid variable capture can be inferred from the
rules.  We prove confluence in Section~\ref{sec:confluence}.

\begin{figure}[!ht]
{\normalsize
$$
\begin{array}{lll}
\B \qquad&\cont{(\lami x A M)}{(\st N l)}\hspace{0.8cm}& \Rew{} \cont{(\cuti A N x
M)}{l}
\end{array}
\hspace{0.3cm}\strut$$
$$
\textsf{\x'} \left\{
\begin{array}{l}
\hspace{1.55cm}
\begin{array}{c@{\hspace{-0.1cm}}l@{\hspace{1.7cm}}l}
\Bi &\qquad\cont M \el&\Rew{} M\\
\Bii &\qquad \cont {(\cont x l)} {l'}&\Rew{} \cont x {(\conc l {l'})}\\
\Biii &\qquad \cont {(\cont M l)} {l'}&\Rew{} \cont M {(\conc l {l'})}\\\\
\Ai&\qquad \conc{(\st M {l'})} {l}&\Rew{} \st M {(\conc {l'} {l})}\\
\Aii&\qquad \conc{\el} {l}&\Rew{} l\\
\Aiii&\qquad \conc{(\conc{l}{l'})} {l''}&\Rew{} \conc{l}
{(\conc{l'}{l''})}\\
\Aiv&\qquad \conc{l} {\el}&\Rew{} l
\end{array}
\\\\
\textsf{\xsubst':}
\left\{
        \begin{array}{c}
                \begin{array}{lll}
                \Ci &\quad \cuti G P y {\lami x A {M}}&\Rew{} \lami x {\cuti G P y {A}}
{\cuti G P y {M}}\\
                \Cii&\quad \cuti G P y {(\cont y l)}&\Rew{} \cont P {\cuti G P y l}\\
                \Ciii&\quad \cuti G P y {(\cont x l)}&\Rew{} \cont x
{\cuti G P y l}\hfill \mbox{ if }x\neq y\\
                \Civ&\quad \cuti G P y {(\cont M l)}&\Rew{} \cont {\cuti G P y M} {\cuti
G P y l}\\
                \Cv&\quad \cuti G P y {\P x A {B}}&\Rew{} \P x {\cuti G P y {A}}
{\cuti G P y {B}}\\
                \Cvi&\quad \cuti G P y s&\Rew{} s\\\\
\Calpha&\quad\cuti A P y {\alpha(M_1,\ldots,M_n)}&\Rew{}\alpha(\cuti A P
y{M_1},\ldots,\cuti A P y{M_n})\\\\
                \Di&\quad \cuti G P y \el &\Rew{} \el\\
                \Dii &\quad \cuti G P y {(\st M l)}&\Rew{} \st {(\cuti G P y {M})}
{(\cuti G P y {l})}\\
                \Diii&\quad \cuti G P y {(\conc l {l'})}&\Rew{} \conc {(\cuti G P y
{l})} {(\cuti G P y {l'})}\\\\
                \Dbeta&\quad\cuti A P y {\beta(M_1,\ldots,M_n)}&\Rew{}\beta(\cuti A P
y{M_1},\ldots,\cuti A P y{M_n})
                \end{array}
        \end{array}
\right.
\end{array}
\right.
$$
\caption{Reduction Rules}\label{fig:red-rules}
}
\end{figure}

We denote by $\Rew{G}$ the contextual closure of the reduction
relation defined by any system $G$ of 
   rewrite rules.\footnote{Via contextual closure, a rewrite rule for terms can thus apply deep inside lists, and vice versa.} The transitive closure of $\Rew{G}$
is denoted by $\Rewplus{G}$, its reflexive and transitive closure is
denoted by $\Rewn{G}$, and its symmetric reflexive and transitive closure
is denoted by $\Rewsn{G}$. The set of strongly normalising elements (those from
which no infinite $\Rew{G}$-reduction sequence starts) is
$\SN{G}$. When not specified, $G$ is assumed to be the system $\B,\x'$
from Fig.~\ref{fig:red-rules}.

We now show that system $\x'$ is terminating. If we add rule $\B$, then
the system fails to be terminating unless we only consider terms that
are typed in a normalising typing system.

We can define an encoding $\foe{\_}$, given in Fig.~\ref{fig:foe}, 
that maps terms and lists into a first-order
syntax given by the following signature:
$$
\{\blob /0, \unname /1,\deuxname/2,\focutname /2, \fosubname
/2\}\cup\{\fotuplename /n\mid n\in\mathbb N\} 
$$
   \noindent which we then equip 
with the well-founded precedence relation defined by
$$
\blob\pre\unname\pre\deuxname\pre\fotuplename[0]\pre\ldots\pre\fotuplename[n]\pre \fotuplename[n+1]\pre\ldots\pre\focutname\pre  \fosubname 
$$

\noindent 
The \emph{lexicographic path ordering} (\textsf{lpo}) induced on the
first-order terms is also well-founded (definitions and results can be
found in~\cite{KL80}, or~\cite[ch. 6]{Terese03}). 

\begin{figure}[!h]
{\normalsize
$$
\begin{array}{|lll|}
\hline
\foe s&&=\blob\\
\foe {\lami x A M}&&=\deux {\foe A}{\foe M}\\
\foe{\P x A M}&&=\deux {\foe A}{\foe M}\\
\foe {\cont x l}&&=\un{\foe l}\\
\foe {\cont M l}&&=\focut{\foe M}{\foe l}\\
\foe {\cuti A M x N}&&=\fosub {\foe M}{\foe N}\\
\foe {\alpha(M_1,\ldots,M_n)}&&=\fotuple[n] {\foe {M_1},\ldots,\foe {M_n}}\\
\hline
\foe {\el}&&=\blob\\
\foe {\st M l}&&=\deux {\foe M}{\foe l}\\
\foe {\conc l {l'}}&&=\deux {\foe l}{\foe {l'}}\\
\foe {\cuti A M x l}&&=\fosub {\foe M}{\foe l}\\
\foe {\beta(M_1,\ldots,M_n)}&&=\fotuple[n] {\foe {M_1},\ldots,\foe {M_n}}\\
\hline
\end{array}
$$
\caption{First-order encoding}
\label{fig:foe}
}
\end{figure}

\begin{thm}\label{th:SNx}\hfill
\begin{enumerate}[$\bullet$]
\item
If $M\Rew{\x'} M'$ then $\foe M\glpo \foe {M'}$.
\item If $l\Rew{\x'}l'$
then $\foe l\glpo \foe {l'}$.
\end{enumerate}
\end{thm}
   \proof 
By simultaneous induction on $M,l$. 
   \qed

\begin{cor}\label{cor:SNx}
System $\x'$ is terminating (on all terms and lists). 
   \qed 
\end{cor}


\section{$\l$-terms and Confluence}\label{sec:confluence}

In this section we define 
   translations 
between the syntax of \PTSCa\ and
that of {\em Pure Type Systems} (\PTS), \ie a variant of
$\lambda$-terms.  Since, in the latter, the only reduction rule (namely,
$\beta$) is confluent, we infer from the 
   translations 
the confluence of
\PTSCa.

We briefly recall the 
   framework of \PTS. Terms 
have the following syntax:
$$t,u,v,T,U,V,\ldots::=x\sep \;s\sep \;\P x T t\sep \;\lami x T t\sep \;t\;u$$
\noindent 
   with an operational semantics given by 
the contextual closure of the $\beta$-reduction rule 
\mbox{$(\lami x v t)\;u\Rew\beta \subst t x u$}, 
in which the substitution is implicit,
\ie is a meta-operation.

Notice now that meta-variables in \PTSCa\ behave like constants of fixed arities during reduction; so it would be natural to reduce the confluence problem of \PTSCa\ to that of a $\lambda$-calculus extended with 
   such constants. 
   We avoid proving confluence of such an extension of \PTS\ with constants. Instead we consider such a constant, say of arity $k$, directly as a free variable applied to (at least) $k$ arguments (indeed, such an approach could also justify confluence for the extended system). 

   Consequently we set aside 
some of the traditional variables of \PTS\ for the specific purpose of encoding meta-variables of \PTSCa: for each meta-variable
$\alpha$ (resp.\ $\beta$) of arity $k$, we reserve in the syntax of \PTS\ a
variable which we write $\alpha^k$ (resp.\ $\beta^k$).

For the remainder of this section, we therefore restrict our attention to that fragment, 
\PTSa, of \PTS-terms where such a variable $\alpha^k$ (\resp $\beta^k$) is never bound and is applied to at least $k$ (\resp $k+1$) arguments. 
   The only subtlety, explained below, is 
why $\beta^k$ is applied to at least $k+1$ arguments (instead of the expected $k$).  

\begin{rem} The fragment 
\PTSa\ is stable under $\beta$-reduction,\footnote{By the capture-avoiding properties of $\beta$-reduction and the fact that, if an occurrence of a free variable is applied to (at least) \(k\) arguments, so are its residuals after a $\beta$-step.} and thus satisfies 
   confluence. 
\end{rem} 


\begin{figure}[!h]
$$
\begin{array}{|lllll|} \upline &\k{\P x A B}&\eqdef&\P x {\k A} {\k B}&\\ &\k{\lami x A M}&\eqdef&\lami x{\k A}{\k M}&\\ &\k{s}&\eqdef&s&\\ &\k{\cont x l}&\eqdef&\subst{\kl z l} z x&\mbox{$z$ fresh}\\ &\k{\cont M l}&\eqdef&\subst {\kl z l} z {\k M}&\mbox{$z$ fresh}\\ &\k{\cuti K P x M}&\eqdef&\subst{\k M}x{\k P}&\\ &\k{\alpha(M_1,\ldots,M_n)}&\eqdef& \alpha^n\ \k{M_1}\ldots\k{M_n}& \midline &\kl y{\el}&\eqdef&y&\\ &\kl y{\st M l}&\eqdef&\subst{\kl z l}z{y\;\k M}&\mbox{$z$ fresh}\\ &\kl y{\conc l {l'}}&\eqdef&\subst{\kl z {l'}}z{\kl y l}&\mbox{$z$ fresh}\\ &\kl y{\cuti v P x l}&\eqdef&\subst{\kl y l}x{\k P}&\\ &\kl y{\beta(M_1,\ldots,M_n)}&\eqdef&\beta^n\ y\ \k {M_1}\ldots\k {M_n}& \downline
\end{array}
$$
\caption{From $\PTSCa$ to $\PTSa$}
\label{fig:PTSCPTS}
\end{figure}

Fig.~\ref{fig:PTSCPTS} shows the 
   translation 
of the syntax of \PTSCa\ into 
   \PTSa. 
While the 
   translation 
of meta-variables for terms is natural, that of  meta-variables for lists is more subtle, since the translation of lists is parameterised by the future head variable. How can we relate such a variable to a list of terms that is (yet) unknown? We simply give it as an extra argument (the first one) of the encoded meta-variable.

\begin{thm}[Simulation of \PTSCa]\label{th:knamesimul}
$\Rew{\beta}$ simulates $\Rew{\B\x'}$ through $\kname$. 
\end{thm}
   \proof 
If $M\!\Rew{\B}N$ then $\k M\Rewn{\beta}\k N$, if
  $l\!\Rew{\B}l'$ then $\kl y l\Rewn{\beta}\kl y {l'}$, 
 if $M\Rew{\x'}N$ then $\k M=\k N$ and if
  $l\Rew{\x'}l'$ then $\kl y l=\kl y {l'}$, which are proved by
simultaneous induction on the derivation step and case analysis.
   \qed 

\begin{figure}[h!]
$$
\begin{array}{|llll|}
\upline
\kCC{s}&\eqdef&s&\\
\kCC{\P x T {U}}&\eqdef&\P x {{\kCC T}}{{\kCC {U}}}&\\
\kCC{\lami x T t}&\eqdef&\lami x {\kCC T}{\kCC t}&\\
\kCC{\alpha^k\ t_1\ldots
  t_k}&\eqdef&\alpha(\kCC{t_1},\ldots,\kCC{t_k})&\\
\kCC{\beta^k\ t\ t_1\ldots
  t_k}&\eqdef&\kCC[\beta(\kCC{t_1},\ldots,\kCC{t_k})] t&\\
\kCC{t}&\eqdef&\kCC[\el]{t}&\mbox{otherwise}
\midline
\kCC[l]{\alpha^k\ t_1\ldots
  t_k}&\eqdef&\cont{\alpha(\kCC{t_1},\ldots,\kCC{t_k})}l&\\
\kCC[l]{\beta^k\ t\ t_1\ldots
  t_k}&\eqdef&\kCC[\conc{\beta(\kCC{t_1},\ldots,\kCC{t_k})}l] t&\\
\kCC[l] {t\;u}&\eqdef&\kCC[\st {{\kCC u}\,} {\, l}] t&\mbox{otherwise}\\
\kCC[l] {x}&\eqdef&\cont x l&\\
\kCC[l] {t}&\eqdef&\cont{\kCC t} l&\mbox{otherwise}
\downline
\end{array}
$$
\caption{From $\PTSa$ to $\PTSCa$}
\label{fig:PTSPTSC}
\end{figure}

Fig.~\ref{fig:PTSPTSC} shows the 
   translation from \PTSa\ into 
\PTSCa.\footnote{Note how we spot the situations which arise from encoded
meta-variables, using the explicitly displayed arity
to identify the arguments. } It is simply the adaptation to
the higher-order case of Prawitz's translation from natural deduction
to sequent calculus~\cite{Prawitz}: the 
   translation \(\kCC{t}\) 
of an application
relies on a list-parameterised version \(\kCC[l]{t}\) of the translation.
   Example~\ref{ex:AB} below illustrates how the definitions in Fig.~\ref{fig:PTSPTSC} and Fig.~\ref{fig:PTSCPTS} expand. 

It is not obvious that the inductive definition of the translation is well-founded. To see this we need the following notion:
\begin{defi}[List-needing terms]
  We say that a $\l$-term $t$ \emph{needs} a list $l$ if the pair
  $(t,l)$ satisfies the following property: if $l=\el$ then $t$ is
  either a variable or an application that is not of the form
  $\alpha^k\ t_1\ldots t_k$.\footnote{Remember that we suppose that $\alpha^k$ is applied to at least \(k\) arguments. }
\end{defi}
The inductive definition of the translation is done by structural induction on 
   the term, subject to the consideration that 
$\kCC[l]{t}$ is defined before $\kCC{t}$ if $t$ needs $l$, and that $\kCC[l]{t}$ is defined after  $\kCC{t}$ if not. The terminology comes from the fact that $t$ needs $l$ if and only if $\kCC[l]{t}$ is \textbf {not} a $\Bi$-redex.


In order to prove confluence, we first need the following results:
\begin{lem}\label{lem:reflPTSC}$ $
\begin{enumerate}[\em(1)]
\item
$\kCC t$ is an $\x'$-normal form.\\
If $l$ is $\x'$-normal and $t$ needs $l$ then $\kCC[l] t$ is $\x'$-normal.
\item
If $l\Rew{\B\x'}l'$ then $\kCC[l]t\Rew{\B\x'}\kCC[l']t$.
\item
$\cont {\kCC[l'] t}
l\Rewn{\x'}\kCC[\conc {l'}l] t$ and $\cont {\kCC t} l\Rewn{\x'}\kCC[l] t$.
\item $\cuti v {\kCC u} x {\kCC
t}\Rewn{\x'}\kCC{\subst t x u}$ and $\cuti v {\kCC u} x {\kCC[l]
t}\Rewn{\x'}\kCC[\cuti v {\kCC u} x l]{\subst t x u}$.
\end{enumerate}
\end{lem}
   \proof 
   Each point 
is obtained by straightforward induction on
$t$. Note that in order to prove point 4 we need rules $\Aiii$ and
$\Aiv$. These are not needed (for simulation of $\beta$-reduction
and for confluence) when only ground terms are concerned.
   \qed 

\begin{thm}[Simulation of \PTS]
\label{Th:PTSPTSC}\strut\\
$\Rew{\B\x'}$ (strongly) simulates $\Rew{\beta}$ through $\kCCname$. 
\end{thm}
   \proof 
If $t\Rew{\beta}u$ then $\kCC t\Rewplus{\B\x'}\kCC u$ and 
$\kCC[l] t\Rewplus{\B\x'}\kCC[l] u$, 
   each 
proved by induction on the derivation step, using
Lemma~\ref{lem:reflPTSC}.4 for the base case and
Lemma~\ref{lem:reflPTSC}.3.
   \qed

Now we study the composition of the two translations:
\begin{lem}\label{lem:comPTSC}Suppose $M$ and $l$ are $\x'$-normal forms.
\begin{enumerate}[\em(1)]
\item
If $t$ needs $l$ then $\kCC[l]{t}=\kCC{\subst {\kl x l} x t}$ (for any $x\notin \FV l$).
\item
$M=\kCC{\k { M}}$.
\end{enumerate}
\end{lem}
   \proof 
By simultaneous induction on $l$ and $M$. Again, rules $\Aiii$ and
$\Aiv$ (as well as $\Calpha$ and $\Dbeta$)  are needed for this lemma to
capture the notion of normal form corresponding to the
$\PTS$-terms, when meta-variables are present.
   \qed

\begin{thm}\label{th:comPTSC}\hfill
\begin{enumerate}[\em(1)]
\item
$\k{\kCC t}=t$
\item
$M\Rewn{\x'}\kCC{\k M}$
\end{enumerate}
\end{thm}
   \proof 
   \hfill 
\begin{enumerate}[(1)]
\item $\k{\kCC t}=t$ and $\k{\kCC[l] t}=\subst {\kl x l} x t$ (with
  $x\neq \FV l$) are obtained by simultaneous induction on $t$.
\item
 $M\Rewn{\x'}\kCC{\k M}$ holds by induction on the longest sequence of
 $\x'$-reduction from $M$ ($\x'$ is terminating): by
 Lemma~\ref{lem:comPTSC}.2, it holds if $M$ is an $\x'$-normal
 form, and if $M\Rew{\x'}N$ then we can apply the induction hypothesis
 on $N$ and by Theorem~\ref{th:knamesimul} we have the result.\qed
\end{enumerate}

\begin{example}\label{ex:AB}Here is an example illustrating Theorem~\ref{th:comPTSC}.1:
$$
\begin{array}{rlllllll}
&&\k{\kCC{\beta^k(x\ y)t_1\ldots t_k}}
&=& \k{\kCC[D]{x\ y}}\\
&=& \k{\cont x {\st{(\cont y\el)} D}}
&=& \kl x {\st{(\cont y\el)} D}\\
&=& \subst{\kl z D}z{x\ \k{\cont y\el}}
&=& \subst{\kl z D}z{x\ \kl y {\el}}\\
&=& \subst{\kl z D}z{x\ y}
&=& \subst{(\beta^k\ z\ \k{A(t_1)} \ldots \k{A(t_k)})}z{x\ y}\\
&=& \beta^k (x\ y) \k{\kCC{t_1}}\ldots\k{\kCC{t_k}} 
\end{array}
$$
where $D = \beta(A(t_1),\ldots,A(t_k))$.
\end{example}

We finally get confluence:

\begin{cor}[Confluence]\label{cor:confluencePTSC}
$\Rew{\x'}$ and $\Rew{\B\x'}$ are confluent. 
\end{cor}

\begin{figure}[!h]
{\normalsize 
$$
\xymatrix{
&&{\quad\ar[dr]^{*}_{\B\x'}} \ar[dl]_{*}^{\B\x'} \ar@{=>}[dd]^{\kname{}}\\
&\quad\ar@/_0.6cm/[dddd]^{*}_{\B\x'}\ar@{=>}[dd]^{\kname{}}&&{\quad}\ar@{=>}[dd]^{\kname{}}\ar@/^0.6cm/[dddd]_{*}^{\B\x'}\\
&&{\quad}\ar[dr]^{*}_{\beta} \ar[dl]_{*}^{\beta}\\
&{\quad}\ar[dr]^{*}_{\beta}\ar@{=>}[dd]^{\kCCname}&&
{\quad}\ar[dl]_{*}^{\beta}\ar@{=>}[dd]^{\kCCname}\\
&&{\quad}\ar@{=>}[dd]^{\kCCname}\\
&{\quad}\ar[dr]^{*}_{\B\x'}& &{\quad}\ar[dl]_{*}^{\B\x'}&\\
&&&
 }
$$
\caption{Confluence by simulation}\label{fig:confbysim}
}
\end{figure}
   \proof 
We use the simulation technique, as for instance
in~\cite{lengrandkesner}: 
consider two reduction sequences starting
from a term in \PTSCa. They can be simulated through $\kename$ by
$\beta$-reductions, and since \PTSa{} is confluent, we can
close the diagram. Now the lower part of the diagram can be simulated
through $\kCCname$ back in \PTSCa, which closes the diagram
there as well, as shown in Fig.~\ref{fig:confbysim} for $\B\x'$.
Notice that
the proof of confluence has nothing to do with typing and does not
rely on any result in Section~\ref{sec:type} (in fact, we use confluence in the
proof of Subject Reduction in the Appendix).
   \qed 

Considering meta-variables in the style of \CRS~\cite{Klo80} avoids
the usual problem of non-confluence coming from the critical pair
between $\B$ and $\Civ$ which generate the two terms $\cuti{} N x
{\cuti{}P y M}$ and $\cuti{}{\cuti{} N x P} y{\cuti{} N x M}$. Indeed,
with \ERS-style meta-variables these two terms need not reduce to a
common term, but with the \CRS-approach, they now can (using the rules
$\Calpha$ and $\Dbeta$). Again, note how the critical pair between
$\Biii$ and itself (or $\Bii$) needs rule $\Aiii$ in order to be
closed, while it was only there for convenience when all terms were
ground.


\section{Typing system and properties}
\label{sec:type}

Throughout this section we consider 
\PTSC, that is, the restriction to \emph{ground} terms of \PTSCa. 
We thus do not need to consider any notion of meta-variable, nor that of any special variable
distinguished among \PTS\ terms, such as those considered in the previous section.

Given the set of sorts $\SCons$, a particular $\nameg$ is specified by
a set $\Sort\subseteq \SCons^2$ and a set $\Rel\subseteq \SCons^3$.
We shall see an example in Section~\ref{sec:SN}. 

\begin{defi}[Typing Environments]\hfill 
\begin{enumerate}[$\bullet$]
\item
A \emph{typing environment} (henceforth simply: `environment', for brevity's sake) 
is a list \(\Gamma\) of pairs taken from $\SVar\times\STerms$, denoted $(x:A)$.
\item
We define the \emph{domain} of an environment and the \emph{application of a
substitution to an environment} as follows:
$$\begin{array}{lll}
\dom {\emptyenv}=\emptyenv&\quad&\dom
{\Gamma,(x:A)}=\dom{\Gamma},x\\
\cuti B P y {(\emptyenv)}=\emptyenv&&
\cuti B P y {(\Gamma,(x:A))}=\cuti B P y {\Gamma},(x:\cuti B P y
A)
\end{array}
$$
\item
It is useful (see Section~\ref{sec:enum}) to define $\dom\Gamma$
as a list, for which the meaning of $x\in\dom\Gamma$ is clear. If
$\mathcal M$ is a \emph{set} of variables, $\mathcal
M\subseteq\dom\Gamma$ means for all $x\in\mathcal M$,
$x\in\dom\Gamma$. Similarly, $\dom\Gamma\cap\dom\Delta$ is the
\emph{set} $\{x\in\SVar\mid x\in\dom\Gamma\wedge x\in\dom\Delta\}$.

We define the following \emph{inclusion relation} between
environments:
\begin{center}
$\Gamma\subenv\Delta$ if for all $(x:A)\in\Gamma$, there is
$(x:B)\in\Delta$ with $A\conv B$. 
\end{center}
\end{enumerate}
\end{defi}

\noindent 
The inference rules in Fig.~\ref{fig:SCoCtyping} inductively define
the derivability of three kinds of statement:
\begin{enumerate}[(1)]
\item $\Gamma\wf$ \\
Intuitively, the derivability of this statement means that the environment $\Gamma$ is well-formed.
\item $\Deri \Gamma M A$ `term typing'\\
Intuitively, the derivability of this statement means that $M$ is of type $A$ in the environment $\Gamma$ (is a proof of $A$ from the assumptions in $\Gamma$).
\item $\Deris \Gamma B l C$ `list typing'\\
The position of $B$ in the sequent is a special place called the \emph{stoup}. 
Intuitively, the derivability of this statement means that, in the environment $\Gamma$, the list $l$ codes for an actual list of terms such that, when something of type $B$ is applied to them, the result is of type $C$ (this codes for a natural deduction of $C$ from $B$ by a series of $\Pi$-elimination rules, whose minor premisses are derived by the proofs-terms in $l$ using the assumptions in $\Gamma$).
\end{enumerate}
Side-conditions are used, such as
$(s_1,s_2,s_3)\in\mathcal R$, $x\not\in\dom\Gamma$, $A\conv B$ or
$\Gamma\subenv\Delta$, and we use the abbreviation
$\Gamma\subenv\Delta\wf$ for $\Gamma\subenv\Delta$ and $\Delta\wf$. We
freely abuse the notation in the customary way, by not distinguishing
between a statement and its derivability according to the rules of
Fig.~\ref{fig:SCoCtyping}. 

There are three conversion rules $\convr$, $\convr'$, and $\convl$ in
order to deal with the two kinds of typing statement and, 
   for list typing, also to be able to 
convert the type in the stoup. 

Because substituting for a variable in
an environment affects the rest of the environment (which could depend
on that variable), the two rules for explicit substitutions ($\cut_2$
and $\cut_4$) must have a particular shape that manipulates the
environment, if the $\nameg$ is to satisfy basic required properties
like those of a \PTS.  

\begin{example}
Here is, as an example, a derivation of $\Deri{x\tpd s_1} {\cont x \el} {s_1}$ in a \PTSC\ where $(s_1,s_2)\in\Sort$.
$$
\infer[{\contr x}]{\Deri{x\tpd s_1} {\cont x \el} {s_1}}
{
  \infer[\ax]{\Deris{x\tpd s_1} {s_1}{\el} {s_1}}
  {
    \infer[\Ttyped]{\Deri{x\tpd s_1} {s_1}{s_2}}
    {
      \infer[\extend]{{x\tpd s_1} \wf}
      {
        \infer[\Ttyped]{\Deri{\emptyenv} {s_1}{s_2}}
        {
          \infer[\empt]{\emptyenv\wf}{}
            \quad ({s_1},{s_2})\in\Sort
        }
      }
            \quad ({s_1},{s_2})\in\Sort
    }
  }
}
$$
\end{example}\medskip

\noindent The lemmas of this section are proved by
straightforward inductions on typing derivations:

\begin{figure}[!t]
{\small 
$$
\begin{array}{|c|}
\hline\\ \infers{}{\emptyenv \wf}\empt \hspace{30 pt}
\infers{{\Deri \Gamma A s}\quad x\notin \dom
\Gamma}{\Gamma,(x\tpd A) \wf}\extend
\\\\
\infers{\Gamma\wf\quad (s,s')\in\Sort}{\Deri \Gamma {s}{s'}}\Ttyped \hspace{30 pt}
\infers{{\Deri \Gamma A {s_1}}\quad {\Deri {\Gamma,(x\tpd A)}
{B}{s_2}}\quad(s_1,s_2,s_3)\in \Rel}{\Deri \Gamma {\P x A {B}} {s_3}}
\Picons
\\\\
\infers{ \Deri \Gamma {\P x A {B}} {s}\quad \Deri {\Gamma,(x\tpd A)} M
{B}}{\Deri \Gamma {\lami x A M} {\P x A B}}\Pir \hspace{20 pt}
\infers{\Deris {\Gamma} A l {B}\quad (x\tpd A)\in \Gamma }{\Deri
\Gamma {\cont x l} {B}}{\contr x} \hspace{30 pt} \infers {\Deri
\Gamma A s}{\Deris{\Gamma} A \el A}\ax
\\\\
\infers{{\Deri \Gamma M A}\quad {\Deri \Gamma {B} s}\quad
A\conv B}{\Deri {\Gamma} M {B}} \convr \hspace{30 pt}
\infers{\Deri \Gamma {\P x A {B}} {s}\quad \Deri \Gamma M
A\quad\Deris {\Gamma}{\cuti A M x B} {l} C}{ \Deris \Gamma {\P x A
{B}} {\st M l} C} \Pil \\\\
\infers{{\Deris \Gamma C l A}\quad {\Deri \Gamma {B} s}\quad
A\conv B }{\Deris {\Gamma} C l {B}}{\convr'}
\hspace{30 pt} 
\infers{{\Deris \Gamma A l C}\quad {\Deri \Gamma {B}
s}\quad A\conv B }{\Deris {\Gamma} B l {C}}{\convl}
\\\\\hline\\
\infers{\Deris
\Gamma C {l'} A\qquad\Deris {\Gamma} A {l} B}{\Deris {\Gamma} C
{\conc {l'} l} {B}}{\cut_1}\hspace{30 pt} 
\infers{\Deri \Gamma P A\qquad\Deris {\Gamma,(x\tpd A),\Delta} B {l} C
\quad \Gamma,\cuti A P x {\Delta}\subenv {\Delta'}\wf}
{ \Deris {{\Delta'}} {\cuti A P x {B}} {\cuti A P x {l}}
{\cuti A P x {C}}}{\cut_2}\\\\
\infers{\Deri \Gamma M A\qquad\Deris {\Gamma} A {l} B}{ \Deri
{\Gamma} {\cont M l} {B}}{\cut_3} \hspace{30 pt} 
\infers{\Deri \Gamma P A\qquad\Deri {\Gamma,(x\tpd A),\Delta} {M}
C\quad \Gamma,\cuti A P x {\Delta}\subenv {\Delta'}\wf}
{\Deri {{\Delta'}}
{\cuti A P x {M}} {C'}}{\cut_4} \\
\hfill\mbox{where either $(C'=C\in\SCons)$ or
$C\not\in\SCons$ and $C'=\cuti A P x {C}$}\\
\\\hline
\end{array}
$$
}
{
\caption{Typing rules of a $\nameg$}\label{fig:SCoCtyping}
}
\end{figure}

\begin{lem}[Properties of typing statements]\label{well-formed}
If \mbox{$\Deri {\Gamma} M A$} (respectively, \mbox{$\Deris {\Gamma} B
  l C$}) then $\FV M\subseteq \dom\Gamma$ (respectively, $\FV
l\subseteq \dom\Gamma$), and the following statements can be derived
with strictly smaller typing derivations:\vfill\eject
\begin{enumerate}[\em(1)]
\item
$\Gamma\wf$
\item
$\Deri \Gamma A s$ for some $s\in\SCons$, or $A\in\SCons$\\
(resp. $\Deri \Gamma B s$ and $\Deri \Gamma C {s'}$ for some $s,s'\in\SCons$)
   \qed
\end{enumerate}
\end{lem}

\begin{cor}[Properties of well-formed environments]$ $\label{wf}
\begin{enumerate}[\em(1)]
\item
If $\Gamma,x:A,\Delta\wf$ then $\Deri\Gamma A s$ for some $s\in\SCons$
with $x\not\in\dom{\Gamma,\Delta}$ and $\FV A\subseteq \dom\Gamma$
(and in particular $x\not\in \FV A$)
\item
If $\Gamma,\Delta\wf$ then $\Gamma\wf$.
   \qed
\end{enumerate}
\end{cor}

\begin{lem}[Weakening]\label{weak1}
Suppose $\Gamma,\Gamma'\wf$ and
$\dom{\Gamma'}\cap\dom\Delta=\emptyset$.
\begin{enumerate}[\em(1)]
\item
If $\Deri {\Gamma,\Delta} M A$ then $\Deri {\Gamma,\Gamma',\Delta}
M A$.
\item
If $\Deris {\Gamma,\Delta} B l C$, then  $\Deris
{\Gamma,\Gamma',\Delta} B l C$.
\item
If ${\Gamma,\Delta}\wf$, then ${\Gamma,\Gamma',\Delta}\wf$.
   \qed
\end{enumerate}
\end{lem}
We can also strengthen the \emph{weakening property} into the
\emph{thinning property} by induction on the typing derivation. This allows
to weaken the environment, permute it, and convert the types inside,
as long as it remains well-formed:
\begin{lem}[Thinning]\label{weak}
Suppose $\Gamma\subenv\Delta\wf$.
\begin{enumerate}[\em(1)]
\item
If $\Deri {\Gamma} M A$ then $\Deri {\Delta} M A$.
\item
If $\Deris {\Gamma} B l C$, then  $\Deris {\Delta}
B l C$.
   \qed 
\end{enumerate}
\end{lem}


Using all of the results above, 
   we obtain Subject Reduction: 
\begin{thm}[Subject Reduction in a \nameg]$ $\label{th:SR}
\begin{enumerate}[\em(1)]
\item
If $\Deri \Gamma M A$ and $M\Rew{} M'$, 
then $\Deri \Gamma {M'} A$
\item
If $\Deris \Gamma B l C$ and $l\redT l'$, 
then $\Deris \Gamma B {l'} C$
\end{enumerate}
\end{thm}
   \proof 
  See the Appendix. 
   \qed 


\section{Correspondence with \PTS}\label{sec:pts}
\subsection{Type preservation}
\label{sec:type-preserve}

There is a logical correspondence between a $\PTSC$ given by the sets
$\SCons$, $\Sort$ and $\Rel$ and the associated \PTS{} given by the same sets.

We prove this by showing that (when restricted to ground terms) the
   translations 
preserve typing.

\begin{figure}[ht]
$$
\begin{array}{|c|}
\hline\\ \infers{}{\emptyenv \wf}{} \hspace{30 pt}
\infers{{\DeriCoC \Gamma T s}\quad x\notin \dom
\Gamma}{\Gamma,(x:T) \wf}{}
\hspace{30 pt}
\infers {\Gamma \wf\quad (x:T)\in\Gamma}{\DeriCoC{\Gamma} x T}{}
\\\\
\infers{\Gamma\wf\quad (s,s')\in\Sort}{\DeriCoC \Gamma s
{s'}}{} \hspace{30 pt} \infers{{\DeriCoC \Gamma U {s_1}}\quad
{\DeriCoC {\Gamma,(x:U)} {T}{s_2}}\quad (s_1,s_2,s_3)\in\Rel}{\DeriCoC
\Gamma {\P x U {T}} {s_3}} {}
\\\\
\infers{ 
\DeriCoC \Gamma {\P x U {T}} {s}\quad 
\DeriCoC{\Gamma,(x:U)} t {T}
}{\DeriCoC \Gamma {\lami x U t} {\P x U T}}{} 
\hspace{20 pt}\infers{\DeriCoC \Gamma t {\P x U {T}}\quad \DeriCoC
\Gamma u U}{ \DeriCoC
\Gamma  {t\;u} {\subst {T} x u}} {}
 \\\\
\hline\\ \infers{{\DeriCoC \Gamma t U}\quad {\DeriCoC \Gamma {V}
s}\quad U\Rewsn{\beta} V}{\DeriCoC {\Gamma} t {V}}
{}\\\\\hline
\end{array}
$$
{\normalsize
\caption{Typing rules of a $\PTS$}
}
\label{fig:CoC}
\end{figure}

Terms in \PTS\ are typed according to the typing rules in Fig.~\ref{fig:CoC}, which
depend on the sets $\SCons$, $\Sort$ and $\Rel$. 
Besides confluence for \(\beta\)-reduction, \PTS{s} have 
the following meta-theoretic properties 
(for proofs, see e.g.~\cite{Barendregt:hlcs1992}):

\begin{thm}\hfill
\begin{enumerate}[\em(1)]
\item
If $\DeriCoC \Gamma t T$ and $\Gamma\subenv\Delta\wf$ then
$\DeriCoC \Delta t T$ (where the relation $\subenv$ is defined
similarly to that of $\nameg$, but with $\beta$-equivalence).
\item
If $\DeriCoC \Gamma t T$ and $\DeriCoC {\Gamma,y:T,\Delta} u {U}$\\
then $\DeriCoC {\Gamma,\subst\Delta y t} {\subst u y t} {\subst {U}
y t}$.
\item
If $\DeriCoC \Gamma t T$ and $t\Rew{\beta} u$ 
then $\DeriCoC \Gamma u T$. 
   \qed 
\end{enumerate}
\end{thm}

We now extend the 
   translations 
to environments:
$$\begin{array}{lll}
\kCC {\emptyenv}=[]&\qquad&\ke {\emptyenv}=[]\\
\kCC {\Gamma,(x:T)}=\kCC {\Gamma},(x:\kCC T)&&\ke {\Gamma,(x:A)}=\ke
{\Gamma},(x:\ke A)
\end{array}$$

Now note that the simulations in Section~\ref{sec:confluence} imply:
\begin{cor}[Equational theories]\hfill\\ 
$t\Rewsn{\beta} u$ if and only if $\kCC t\Rewsn{\Herb}\kCC u$\\
$M\Rewsn{\All} N$ if and only if $\ke M\Rewsn{\beta}\ke N$ 
\qed 
\end{cor}

Preservation of typing is proved by induction on the typing derivations:
\begin{thm}[Preservation of typing 1]
\label{Th:typePTSPTSC}\hfill
\begin{enumerate}[\em(1)]
\item
If $\DeriCoC\Gamma t T$ then $\Deri{\kCC \Gamma}{\kCC t}{\kCC T}$
\item
If $(\DeriCoC\Gamma {t_i}{\subst {\cdots\subst
{T_i}{x_{1}}{t_{1}}}{x_{i-1}}{t_{i-1}}})_{i=1\ldots n}$\\
and
 $\Deri{\kCC \Gamma}{\kCC{\P {x_1}{T_1}{\ldots\P {x_n}{T_n}{T}}}}s$\\
then $\Deris{\kCC \Gamma}{\kCC{\P {x_1}{T_1}{\ldots\P
{x_n}{T_n}{T}}}} {\kCCC{t_1\ldots t_n}}{\kCC {\subst {\cdots\subst
{{T}}{x_{1}}{t_{1}}}{x_{n}}{t_{n}}}}$
\item
If $\Gamma \wf$ then $\kCC \Gamma\wf$ 
   \qed 
\end{enumerate}
\end{thm}

\begin{thm}[Preservation of typing 2]
\label{Th:typePTSCPTS}\hfill
\begin{enumerate}[\em(1)]
\item
If $\Deri\Gamma M A$ then $\DeriCoC{\ke \Gamma}{\ke M}{\ke A}$
\item
If $\Deris\Gamma B l C$ then $\DeriCoC{\ke \Gamma,y:\ke B}{\kel y l}{\ke C}$ for any fresh $y$
\item
If $\Gamma \wf$ then $\ke \Gamma\wf$
   \qed 
\end{enumerate}
\end{thm}


\subsection{Equivalence of Strong Normalisation}
\label{sec:SN}
\begin{thm}
A $\nameg$ given by the sets $\SCons$, $\Sort$, and $\Rel$ is strongly
normalising if and only if the corresponding \PTS{} given by the same sets is.
\end{thm}
   \proof
Assume that the \nameg{} is strongly normalising, and let us consider
a well-typed $t$ of the corresponding \PTS, i.e. $\DeriCoC\Gamma t T$
for some $\Gamma,T$. By Theorem~\ref{Th:typePTSPTSC}, $\Deri{\kCC
\Gamma}{\kCC t}{\kCC T}$ so $\kCC t\in\SN{}$. Now by
Theorem~\ref{Th:PTSPTSC}, any reduction sequence starting from $t$ maps
to a reduction sequence of at least the same length starting from
$\kCC t$, but those are finite.

Now assume that the \PTS{} is strongly normalising and that
$\Deri\Gamma M A$ in the corresponding $\nameg$. By subject reduction,
any $N$ such that $M\Rewn{}N$ satisfies $\Deri\Gamma N A$ and any
sub-term $P$ (resp. sub-list $l$) of any such $N$ is also typable.  By
Theorem~\ref{Th:typePTSCPTS}, for any such $P$ (resp. $l$), $\ke P$
(resp. $\kel y l$) is typable in the
\PTS, so it is strongly normalising by assumption.

We now refine the first-order encoding of any such $P$ and $l$ (as
defined in Section~\ref{sec:synsem}), emulating the technique of Bloo
and Geuvers~\cite{BlooGeuvers}.

Accordingly, we refine the first-order signature from Section~\ref{sec:synsem}
by \emph{labelling} the symbols $\focut [t]\_\_$ and $\fosub [t] \_\_
$ with all strongly normalising terms $t$ of a \PTS, thus generating
an infinite signature.  The precedence relation is refined as follows
$$
\blob\pre\un \_\pre\deux \_ \_\pre\focut [t]
\_\_\pre \fosub [t] \_\_ 
$$
but we also set $\fosub [t]\_\_\pre\focut
[t']\_\_$ whenever $t'\Rewplus{\beta}t$. The precedence is still well-founded,
so the induced (\textsf{lpo}) is also still well-founded (definitions and results can be
found in~\cite{KL80}). The refinement of the encoding is given in Fig~\ref{fig:fot}.
An induction on terms shows that reductions decrease the
\textsf{lpo}.
   \qed 

\begin{figure}[!h]
{\normalsize
$$
\begin{array}{|lll|}
\hline
\fot s&&=\blob\\
\fot {\lami x A M}&=\fot{\P x A M}&=\deux {\fot A}{\fot M}\\
\fot {\cont x l}&&=\un{\fot l}\\
\fot {\cont M l}&&=\focut[\ke {\cont M l}]{\fot M}{\fot l}\\
\fot {\cuti A M x N}&&=\fosub[\ke {\cuti A M x N}] {\fot M}{\fot N}\\
\fot {\el}&&=\blob\\
\fot {\st M l}&&=\deux {\fot M}{\fot l}\\
\fot {\conc l {l'}}&&=\deux {\fot l}{\fot {l'}}\\
\fot {\cuti A 
M x N}&&=\fosub[\ke {\cuti A M x l}] {\fot M}{\fot l}\\
\hline
\end{array}
$$
\caption{First-order encoding}
\label{fig:fot}
}
\end{figure}

Examples of strongly normalising \PTS{} are the \emph{Calculus of 
Constructions}~\cite{CoC}, on which the proof-assistant
\Coq\ is based~\cite{Coq} (but it also uses inductive types and local
definitions), as well as the other systems of Barendregt's Cube,
for all of which we now have a corresponding \nameg{} that can be used
for proof-search.

\ignore{
Let $\SCons=\{\K_0,\K_1,\ldots,\K_i,\ldots\}$.
\ignore{We sometimes write $\T$ for $\K_0$.}
Let $\Sort=\{(\K_i,\K_{i+1})|\;i\in\mathbb N\}$ and\linebreak
$\Rel=\{(\K_i,\K_j,\K_{max (i,j)})|\;i,j\in\mathbb N\}\cup
\{(\K_i,\K_0)|\;i\in\mathbb N\}$.

$\CoC$ contains as a sub-system all of Barendregt's Cube: the
simply-typed $\l$-calculus, system $F$, system $F_\omega$ (which can
be seen as a type-assignment system for Higher-Order Logic), and their
respective versions with dependent types.  The Calculus of
Constructions is also the theoretical basis of the proof-assistant
\Coq~\cite{Coq} (which also uses inductive types and local
definitions).

We call \emph{Sequent Calculus of Constructions with Universes}
($\name$) the particular
\nameg{} built with the same sets $\SCons$, $\Sort$ and $\Rel$ as in $\CoC$.
The strong normalisation of $\CoC$ implies the strong normalisation
of all its sub-systems such as those of Barendregt's Cube, for all of
which we now have a sequent calculus version as a \nameg.
In some of them, proof-search can be simplified or made more deterministic.
}


\section{Proof-search}
\label{sec:proofsearch}

Proof-search considers as inputs an environment $\Gamma$ and a
type $A$, and the output, if successful, will be a
term $M$ such that $\Deri \Gamma M A$, moreover one in normal form.
When we search for a list 
   \(l\) such that 
$\Deris \Gamma B l C$, the type $B$ in the
stoup is also an input. Henceforth, 
   such a term type $A$ or list type $C$ 
will be called simply a \emph{goal}.

The inference rules now need to be \emph{syntax-directed}, that is
determined by the shape of the goal (or of the type in the stoup), and
the proof-search system (\PSS, for short) is then obtained by
optimising appeals to the conversion rules, yielding the presentation
given in Fig.~\ref{fig:Proof-search}.  The incorporation of the
conversion rules into the other rules is similar to that of the
Constructive Engine in natural
deduction~\cite{Hue89,jutting-mckinna-pollack}; however that algorithm was
designed for type synthesis, for which the inputs and outputs are not
the same as in proof-search, as mentioned in the introduction.

\begin{figure}[!ht]
$$
\begin{array}{|c|}
\hline\\
\infers { D\conv C}{\DeriPSs{\Gamma} D \el {C}}{\ax}
\quad
\infers{D\Rewn{}\P x A B\quad\DeriPS \Gamma M A\quad\DeriPSs
{\Gamma}{\cuti A M x B} {l} C}{ \DeriPSs \Gamma {D} {\st M l} C} {\Pil}
\\\\
\hline\\
\infers{{C\Rewn{\Herb}{s_3}\quad (s_1,s_2,s_3)\in R\quad\DeriPS \Gamma A {s_1}}\quad {\DeriPS {\Gamma,(x:A)}
{B}{s_2}}}{\DeriPS
\Gamma {\P x A {B}} {C}} {\Picons}
\\\\
\infers{C\Rewn{}s'\quad (s,s')\in\Sort}{\DeriPS \Gamma s C}{\Ttyped}\qquad
\infers{ (x:A)\in \Gamma\quad\DeriPSs {\Gamma} A l {C}}{\DeriPS \Gamma
{\cont x l} {C}}{\contr x}
\\\\
\infers{C\Rewn{\Herb}\P x A
B \quad \DeriPS {\Gamma,(x:A)} M {B}}{\DeriPS \Gamma {\lami x A M} {C}}{\Pir}
\qquad
 \\\\
\hline
\end{array}
$$
\caption{Rules for Proof-search}\label{fig:Proof-search}
\end{figure}

\noindent Note one small difference from~\cite{LDMcK06}: we do not, in rule
$\Pir$, require that $A$ be a normal form.  As in \cite{LDMcK06},
soundness and completeness hold, but because of this difference, we
get \emph{quasi-normal forms} rather than normal forms.

\begin{defi}[Quasi-normal form]
A term (or a list) is a \emph{quasi-normal form} if all its redexes
are within type annotations of $\l$-abstractions, \eg $A$ in $\lami x
A M$.
\end{defi}

Notice that, as we are searching for (quasi-)normal forms, there are
\emph{no} cut-rules  in \PSS{}.  However, in \PTSC\ even terms in
normal form may need instances of the cut-rule in their typing
derivation. 
This is because, in contrast to logics where well-formedness of formulae is
pre-supposed (such as first-order logic, where cut is admissible),
\PTSC\ checks well-formedness of types. For instance in rule $\Pil$ of
\PTSC\, a type which is not
normalised ($\cuti A M x B$) occurs in the stoup of the third
premiss, so cuts might be needed to type it inside the
derivation.

We conjecture that if we modify rule $\Pil$ by now requiring in the
stoup of its third premiss \emph{a normal form} to which $\cuti A M x B$
reduces, then any typable normal form can be typed with a cut-free
derivation. 
However, this would make rule $\Pil$ more complicated and, more
importantly, we do not need such a conjecture to hold in order to
perform proof-search. 

In contrast, system \PSS\ avoids this problem by obviating such type-checking
constraints altogether, because types are the
input of proof-search, and should therefore be checked before starting 
search. This is the spirit of the
type-checking proviso in the following soundness theorem. 

\PSS{} is sound and complete in the following sense:\vfill\eject

\begin{thm}\hfill
\begin{enumerate}[\em(1)]
\item(Soundness) Provided $\Deri\Gamma A s$,
if $\DeriPS \Gamma M A$ then $\Deri \Gamma M A$ and $M$ is a quasi-normal form.
\item(Completeness)
If $\Deri \Gamma M A$ and $M$ is a quasi-normal form, then 
   we can derive 
$\DeriPS \Gamma M A$.
\end{enumerate}
\end{thm}
   \proof
Both proofs are done by induction on typing derivations, with similar
statements for list typing.
For Soundness, the type-checking proviso is verified every time we need
the induction hypothesis. For Completeness, the following lemma is
required (and also proved inductively): 
given $A\conv {\!A'}$, $B\conv {\!B'}$ and $C\conv {\!C'}$,
if $\DeriPS \Gamma M A$ then $\DeriPS \Gamma M {A'}$, and 
if $\DeriPSs \Gamma B l C$ then $\DeriPSs \Gamma {B'} l {C'}$.
   \qed 

Note that neither part of the theorem relies on the unsolved problem
of \emph{expansion
postponement}~\cite{jutting-mckinna-pollack,Poll:EP98}. Indeed, as
indicated above \PSS{}
\emph{does not} check types.
When recovering a full derivation tree from a \PSS{} one by the
soundness theorem, expansions and cuts might be introduced at any point, arising from the
derivation of the type-checking proviso.
\ignore{
Expansions need not be postponed (resp. cuts might be present) in a
full derivation tree recovered from a \PSS{} one, were it only for
those expansions (resp. those cuts) used to derive the type-checking
proviso that the soundness result combines to the \PSS{}-derivation.
}

Basic proof-search can be done in \PSS\ simply by
\begin{enumerate}[$\bullet$]
\item
reducing the goal, or the type in the stoup;
\item
depending on its shape, trying to apply one of the inference rules
bottom-up; and
\item
recursively calling the process on the new goals (called
\emph{sub-goals}) corresponding to each premiss.
\end{enumerate} 

\noindent However, some degree of non-determinism is to be expected in 
proof-search. Such non-determinism is already present in natural
deduction, but the sequent calculus version conveniently identifies
where it occurs exactly.

There are three potential sources of such non-determinism:
\begin{enumerate}[$\bullet$]
\item
The choice of a variable $x$ for applying rule $\contr x$, knowing
only $\Gamma$ and $B$ (this corresponds in natural deduction to the
choice of the head-variable of the proof-term). Not every variable of
the environment will work, since the type in the stoup will eventually have to be
unified with the goal, so we
still need backtracking.
\item
When the goal reduces to a $\Pi$-type, there is an overlap between
rules $\Pir$ and $\contr x$; similarly, when
the type in the stoup reduces to a $\Pi$-type, there is an overlap
between rules $\Pil$ and $\ax$. Both overlaps
disappear when $\contr x$ is restricted to the case when the goal does
not reduce to a $\Pi$-type (and sequents with stoups never have a
goal reducing to a $\Pi$-type).
This corresponds to looking only for $\eta$-long normal forms in
natural deduction. This restriction also brings the derivations in
\LJT{} (and in our \PTSC) closer to the notion of uniform proofs.
Further work includes the addition of $\eta$ to the
notion of conversion in $\PTSC$.
\item
When the goal reduces to a sort $s$, three rules can be
applied (in contrast to the first two points, this source of
non-determinism does not already appear in the
propositional case).
\end{enumerate}
Such classification is often called \emph{``don't care''}
non-determinism in the case of the choice to apply an invertible rule and
\emph{``don't know''} non-determinism
when the choice identifies a potential backtracking point. 

Don't know non-determinism can be in fact 
quite constrained by the need to eventually unify the 
stoup with the goal, as an example in Section~\ref{sec:example} below
illustrates.  
Indeed, the dependency created by a $\Pi$-type forces the searches for
proofs of the two premisses of rule $\Pil$ to be sequentialised in a
way that might prove inefficient: the proof-term produced for the
first premiss, selected among others at random, might well lead to the
failure to solve the second premiss, leading to endless backtracking.

Hence, there is much to be gained by postponing the search for a
proof of the first premiss and trying to solve the second with
incomplete inputs.
This might not terminate with success or
failure but will send back constraints that may be useful in helping
to solve the first premiss with the correct proof-term. ``Helping''
could just be giving some information to orient and speed-up the
search for the right proof-term, but it could well define it completely
(saving numerous attempts with proof-terms that will lead to failure).
Unsurprisingly, these constraints are produced by the axiom rule as
\emph{unification} constraints.

In \Coq~\cite{Coq}, the proof-search tactic \verb"apply x" can be
decomposed into the bottom-up application of $\contr x$ followed by a
series of bottom-up applications of $\Pil$ and finally $\ax$, but it
either postpones the solution of sub-goals or automatically solves
them from the unification attempt, often avoiding obvious
back-tracking.

In the next section we use the framework with meta-variables we have
introduced to capture this behaviour in an extended sequent calculus.


\section{Using meta-variables for proof-search}
\label{sec:enum}

We now use the meta-variables in \PTSCa\ to delay the solution of
sub-goals created by the application of rules such as
\(\Pil\). In this way, the extension from \PTSC\ to \PTSCa\ supports
not only an account of tactics such as \verb"apply x" of \Coq, but
also the specification of algorithms for type inhabitant enumeration
and unification. It provides the search-trees that such algorithms
have to explore.  Our approach has two main novelties in comparison
with similar approaches (in the setting of natural deduction) by
Dowek~\cite{Doweksynthesis} and Mu\~noz~\cite{Munozsynthesis}.

The first main novelty is that the search-tree is made of the inference
rules of sequent calculus and its exploration is merely the root-first
construction of a derivation tree; this greatly simplifies the
understanding and the description of what such algorithms do.

The second main novelty is the avoidance of the complex phenomenon
known as \emph{r-splitting} that features in traditional inhabitation
and unification algorithms (\eg \cite{Doweksynthesis}).  In natural 
deduction, lists of arguments are not first-class objects; hence, when
choosing a head variable in the construction of a $\lambda$-term, one
also has to anticipate how many arguments it will be applied to (with
polymorphism, there could be infinitely many choices). This
anticipation can require a complex analysis of the sorting relations
during a single search step and result in an infinitely branching
search-tree whose exploration requires interleaving techniques. This
is avoided by the use of meta-variables for lists of unknown length,
which allows the choice of a head variable without commitment to the
number of its arguments.

In contrast to Section~\ref{sec:pts}, where we confined our attention
to the \emph{ground} terms of \PTSCa\ and their relation to the
corresponding \PTS, here we consider the full language of \emph{open}
terms, representing incomplete proofs and partially solved goals.
Correspondingly, \emph{(open) environments} are now lists of pairs,
denoted $(x:A)$, where $x$ is a variable and $A$ is a (possibly open)
term (while \emph{ground environments} only feature ground
terms). Ground terms and environments are the eventual targets of
successful proof-search, with all meta-variables instantiated.  We
further consider a new environment $\Sigma$ that contains the
sub-goals that remain to be proved:\vfill\eject

\begin{defi}[Goal environment, constraint, solved constraint, 
substitution]\hfill
\begin{enumerate}[$\bullet$]
\item
A \emph{goal environment} $\Sigma$ is a list of:
\begin{enumerate}[$-$] 
\item Triples of the form $\Deri\Gam \alpha A$, declaring the
meta-variable $\alpha$ and called
\emph{(term-)goals}, where $A$ is an open term and $\Gam$ is an open
environment.  
\item 4-tuples of the form $\Deris\Gam B \beta A$,  declaring the
meta-variable $\beta$ and called
\emph{(list-)goals}, where $A$ and $B$ are open terms and $\Gam$ is an open
environment.  
\item Triples of the form $\constr\Gam A B$, called
\emph{constraints}, where $\Gam$ is an open environment
and $A$ and $B$ are open terms.
\end{enumerate}
Goals of a goal environment are required to declare distinct meta-variables.
\item
A constraint is \emph{solved} if it is of the form
$\constr\Gam A B$ where
$A$ and $B$ are ground and $A\conv B$.
\item
A goal environment is \emph{solved} if it
contains no term or list goals and consists only of solved constraints.
\item 
A \emph{substitution} is a finite function \(\sigma\) that maps a meta-variable
for term (\resp list), of arity $n$, to a closed \emph{higher-order} term
(\resp list) of arity $n$, that is to say, a term (\resp list) under a
series of $n$ bindings that capture (at least) its free variables (\eg
$x.y.M$ with $\FV M\subseteq\{x,y\}$).\footnote{This uses a standard notation that can be found in \eg \cite{Terese03}, Ch.\ 11. }

Such a series of bindings can be provided by a typing environment
$\Gamma$, \eg $\dom\Gamma.M$ (which is a useful notation when \eg
$\Deri \Gamma M A$). 

As usual, substitutions \(\sigma\) are built up from individual
bindings of the form $(\alpha\mapsto x_1\ldots x_n.M)$ by
concatenation \(\sigma, \sigma'\), where bindings in \(\sigma'\)
override those in \(\sigma\). 

\item
The application of a substitution to terms and lists is defined by
induction on these. Only the base cases are interesting: 

If $\sigma(\alpha)=x_1\ldots x_n.M$, then
$\sigma(\alpha(N_1,\ldots,N_n))$ is the $\x'$-normal
form\footnote{Which exists because $\x'$ is convergent even on untyped terms, by
   Corollary~\ref{cor:SNx}. } 
of
$$\cuti{}{\sigma(N_1)}{x_1}{ \ldots
\cuti{}{\sigma(N_n)}{x_n}{M}}$$
(with the usual capture-avoiding conditions).

Similarly, if $\sigma(\beta)=x_1\ldots x_n.l$, then
$\sigma(\beta(N_1,\ldots,N_n))$ is the $\x'$-normal form of
$$\cuti{}{\sigma(N_1)}{x_1}{ \ldots
\cuti{}{\sigma(N_n)}{x_n}{l}}$$
The application of a substitution to an environment is the straightforward
extension of the above.
\end{enumerate}
\end{defi}

\noindent For instance on the example of Section~\ref{subsec:syn}, for an {\em
actual} term $M$ with\linebreak $\FV M=\{x,y\}$ and $\sigma(\alpha)=x.y.M$, we
have that $\sigma(\alpha(N,P))$ is the $\x'$-normal form of
$\cuti{}{\sigma(N)}{x}{\cuti{}{\sigma(P)}{y}{M}}$.

The reason why we $\x'$-normalise the instantiation of meta-variables
is that if $M$ is already $\x'$-normal then $(\alpha\mapsto x_1\ldots
x_n.M)(\alpha(\cont{y_1}\el,\ldots,\cont{y_n}\el))$ really
\emph{is} a renaming of $M$ (and also an $\x'$-normal form).
This ensures that only normal forms are output by our system for
proof-search, which we can more easily relate to \PSS.

\begin{figure}[!h]
{\small
\renewcommand{\TtypedPE}{\Ttyped}
\renewcommand{\axPE}{\ax}
\renewcommand{\PiconsPE}{\Picons}
\renewcommand{\PirPE}{\Pir}
\renewcommand{\PilPE}{\Pil}

\newcommand{\ClaimPE}[1][\alpha]{\ensuremath{\textsf{Claim}_{#1}}}
\newcommand{\ClaimsPE}[1][\beta]{\ensuremath{\textsf{Claim}_{#1}}}
\newcommand{\SolvePE}[1][\alpha]{\ensuremath{\textsf{Solve}_{#1}}}
\newcommand{\SolvesPE}[1][\beta]{\ensuremath{\textsf{Solve}_{#1}}}
\newcommand{\SolvedPE}{\ensuremath{\textsf{Solved}}}
\newcommand{\contrPE}[1][x]{\contr{#1}}
$$
\begin{array}{|c|}
\upline\\[-4pt]
\infer{\enumPSs \Sigma \Gam D C {\beta(\cont {x_1}\el,\ldots,\cont
{x_n}\el)}
{(\Deris \Gam D \beta C)}}
{\Gam=x_1\col A_1,\ldots,x_n\col A_n}
{\ClaimsPE}
\\
\infer{\enumPSs \Sigma \Gam D C \el {\constr \Gam D C}}
{\strut}
{\axPE}
\\\\
\infer{\enumPSs \Sigma \Gam D C {\st M l}{\Sigma_1,\Sigma_2}}
{D\Rewn{\B\x}\P x A B\quad\enumPS\Sigma \Gam A M {\Sigma_1}
\quad \enumPSs {\Sigma,\Sigma_1} \Gam {\cuti A M x B} C {l}{\Sigma_2}}
{\PilPE}
\\[-4pt]\mmidline\\[-4pt]
\infer{\enumPS \Sigma \Gam C {\alpha(\cont {x_1}\el,\ldots,\cont
{x_n}\el)}
{(\Deri \Gam \alpha C)}}
{\Gam=x_1\col A_1,\ldots,x_n\col A_n}
{\ClaimPE}
\\\\
\infer{\enumPS \Sigma \Gam C {s'} \emptyenv}{C\Rewn{\B\x}s\quad(s',s)\in\Sort}
{\TtypedPE}
\\\\
\infer{\enumPS \Sigma \Gam C {\P x A B} {\Sigma_1,\Sigma_2}}
{C\Rewn{\B\x}s\quad(s_1,s_2,s)\in\Rel\quad\enumPS\Sigma \Gam {s_1} A {\Sigma_1}\quad\enumPS\Sigma
{\Gam,x\col A} {s_2} B {\Sigma_2}}
{\PiconsPE}
\\\\
\infer{\enumPS \Sigma \Gam C {\cont x l} {\Sigma'}}
{(x\col A)\in\Gam\quad\enumPSs \Sigma \Gam A C {l} {\Sigma'}}
{\contrPE}
\\\\
\infer{\enumPS \Sigma \Gam C {\lami x A M} {\Sigma'}}
{C\Rewn{\B\x}\P x A B\quad
\enumPS \Sigma {\Gam,x\col A} B {M} {\Sigma'}}
{\PirPE}
\\[-4pt]\mmidline\\[-4pt]
\infer{\goalsPS {\Sigma,(\Deris\Gam B\beta
    C),\Sigma'}{\sigma_\Sigma,(\beta\mapsto \dom\Gamma.(\sigma_{\Sigma},\sigma_{\Sigma''})(l)),\sigma_{\Sigma'}}}
{\enumPSs \Sigma \Gam B C l {\Sigma''}\qquad \goalsPS{\Sigma,\Sigma'',(\beta\mapsto\dom\Gamma.l)
{(\Sigma')}}{\sigma_\Sigma,\sigma_{\Sigma''},\sigma_{\Sigma'}}}
{\SolvesPE}
\\\\
\infer{\goalsPS {\Sigma,(\Deri\Gam\alpha A),\Sigma'}{\sigma_\Sigma,(\alpha\mapsto
\dom\Gamma.(\sigma_{\Sigma},\sigma_{\Sigma''})(M)),\sigma_{\Sigma'}}}
{\enumPS \Sigma \Gam A M {\Sigma''}\qquad
\goalsPS{\Sigma,\Sigma'',(\alpha\mapsto \dom\Gamma.M)
{(\Sigma')}}{\sigma_\Sigma,\sigma_{\Sigma''},\sigma_{\Sigma'}}}
{\SolvePE}
\\\\
\infer{\goalsPS \Sigma\es}{\Sigma \mbox{ is solved}}
{\SolvedPE}
\\[-4pt]\downline
\end{array}
$$
}
\caption{Proof-term enumeration $\seqf[{\PE}]{}$}
\label{fig:inhabitant}
\end{figure}

We now introduce this system, called \PE\ for \emph{Proof Enumeration}, which
can be seen as an extension of \PSS\ to open terms.

\begin{defi}[An inference system \(\PE\) for proof enumeration]\strut\\
The inference rules for system \(\PE\), in Fig.~\ref{fig:inhabitant}, manipulate three kinds of statement:
\begin{enumerate}[$\bullet$]
\item
The first two are of the form $\enum \Sigma \Gam A M {\Sigma}$ and
$\enums \Sigma \Gam B C l {\Sigma}$.
\item
The third kind of statement is of the form $\goals \Sigma \sigma$, where
\begin{enumerate}[$-$]
\item
$\Sigma$ is a goal environment;  
\item
$\sigma$ is a substitution as defined above. 
\end{enumerate}
\end{enumerate}

In the bottom part of the figure we use the notational convention that
a substitution denoted $\sigma_\Sigma$ has the meta-variables of the
goal environment $\Sigma$ as its domain.

Derivability in \PE\ of the three kinds of statement is denoted
   respectively by 
$\enumPS \Sigma \Gam A M {\Sigma}$, 
$\enumPSs \Sigma \Gam B C l {\Sigma}$ 
and $\goalsPS \Sigma \sigma$.
\end{defi}

The statements $\enum \Sigma \Gam A M {\Sigma}$ and $\enums \Sigma \Gam
B C l {\Sigma}$ have the same intuitive meaning as the corresponding
statements in system \PSS, but note the extra goal environment
$\Sigma$, which represents the list of sub-goals and constraints that
have been produced by proof-search and that remain to be solved. Thus,
the inputs of proof enumeration are $\Gam$ and $A$ (and $\Gam,B$ and $C$ for the
second kind of statement) and the outputs are a term $M$ (or list $l$) and
goal environment $\Sigma$.  Statements of \PSS\ are in fact particular cases of these
statements with $\Sigma$ being always solved.

In contrast, in a statement of the form $\goals \Sigma \sigma$,
$\Sigma$ is the list of goals to solve, together with the constraints
that the solutions must satisfy. It is the input of proof enumeration 
and $\sigma$ is meant to be its solution, \ie the output.

Now we prove that \PE\ is \emph{sound}.
For that we need the following notion:
\begin{defi}[Solution]
We define the property $\sigma$ \emph{is a solution of a goal
environment $\Sigma$}, by induction on the length of $\Sigma$.
\begin{enumerate}[$\bullet$]
\item
$\sigma$ is a solution of $\emptyenv$.
\item
If $\sigma$ is a solution of $\Sigma$ and 
$$\DeriPS {x_1\col \sigma(A_1),\ldots,x_n\col \sigma(A_n)}
{(\sigma (\alpha))(\cont{x_1}\el,\ldots,\cont{x_n}\el)}{\sigma(C)}$$
then
$\sigma$ is a solution of $\Sigma,(\Deri {x_1\col A_1,\ldots,x_n\col
  A_n} \alpha C)$.
\item
If $\sigma$ is a solution of $\Sigma$ and 
$$\DeriPSs {x_1\col \sigma(A_1),\ldots,x_n\col \sigma(A_n)}{\sigma(D)}
{(\sigma (\beta))(\cont{x_1}\el,\ldots,\cont{x_n}\el)}{\sigma(C)}$$
then
$\sigma$ is a solution of $\Sigma,(\Deris {x_1\col A_1,\ldots,x_n\col
  A_n} D\beta C)$.
\item
If $\sigma$ is a solution of $\Sigma$ and 
$$
{\sigma(D)}\conv{\sigma(C)}
$$
then
$\sigma$ is a solution of $\Sigma,\constr \Gam D C$.
\end{enumerate}
\end{defi}

\noindent For soundness we also need the following lemma:
\begin{lem}Suppose that $\sigma(M)$ and $\sigma(l)$ are ground.
\begin{enumerate}[\em(1)]
\item
If $M\Rew{\B\x'}N$ then $\sigma(M) \Rewn{\B\x}\sigma(N)$.
\item
If $l\Rew{\B\x'}l'$ then $\sigma(l) \Rewn{\B\x}\sigma(l')$.
\end{enumerate}
\end{lem}
   \proof 
By simultaneous induction on the derivation of the reduction step,
checking all rules for the base case of root reduction.
   \qed 

\begin{thm}[Soundness]\label{th:PEsound}
Suppose $\sigma$ is a solution of $\Sigma$.
\begin{enumerate}[\em(1)]
\item
If $\enumPS \Sigma \Gam A M
{\Sigma}$ then 
$\DeriPS {\sigma(\Gam)} {\sigma(M)} {\sigma(A)}$.
\item
If $\enumPSs \Sigma \Gam B C l
{\Sigma}$ then 
$\DeriPSs {\sigma(\Gam)}{\sigma(B)} {\sigma(l)} {\sigma(C)}$.
\end{enumerate}
\end{thm}
   \proof 
By induction on derivations.
   \qed 
\begin{cor}
If $\goalsPS \Sigma\sigma$ then $\sigma$ is a solution of $\Sigma$.
\end{cor}
   \proof
By induction on the derivation, using Theorem~\ref{th:PEsound}.
   \qed 

System \PE\ is \emph{complete} in the following sense:

\begin{thm}[Completeness]\hfill
\begin{enumerate}[\em(1)]
\item
If $\DeriPS \Gam M A$ then $\enumPS \Sigma \Gam A M {\Sigma}$ 
for some solved $\Sigma$.
\item
If $\DeriPSs \Gam B l C$ then $\enumPSs \Sigma \Gam B C l {\Sigma}$ 
for some solved $\Sigma$.
\end{enumerate}
\end{thm}
   \proof
By induction on derivations. The rules of \PE\ generalise those of \PSS.
   \qed 

\noindent In fact, completeness of the full system \PE\ is not surprising,
since it is quite general. In particular, nothing is said about when
the process should decide to abandon the current goal and start
working on another one. Hence we should be interested in completeness
of particular \emph{strategies} dealing with that question. For instance: 
\begin{enumerate}[$\bullet$]
\item
  We can view the system \PSS\ as supporting the strategy of eagerly
  solving sub-goals as soon as they are created, never delaying them
  with the sub-goal environment.

\item
  The algorithm for proof enumeration in~\cite{Doweksynthesis} would
  correspond here to the ``lazy'' strategy that always abandons the
  sub-goal generated by rule $\PilPS$, but this in fact enables
  unification constraints to guide the solution of this sub-goal
  later, so in that case laziness is probably more efficient than
  eagerness. This is probably what should be chosen for automated
  theorem proving.

\item 
  Mixtures of the two strategies can also be considered and could be
  the basis of interactive theorem proving. Indeed in some cases the
  user's input might be more efficient than the automated algorithm,
  and rule $\PilPS$ would be a good place to ask whether the user has
  any clue to solve the sub-goal (since it could help solving the rest
  of the unification). If he or she has none, then by default the
  algorithm might abandon the sub-goal and leave it for later.

In \Coq, the tactic \verb"apply x" does something similar: it tries to
automatically solve the sub-goals that interfere with the unification
constraint (leaving the other ones for later, visible to the user),
but, if unification fails, it is always possible for the user to
use the tactic and give explicitly the proof-term to make it
work. However, such an input is not provided in proof synthesis mode
   in \Coq\ 
and the user really has to give it fully, since the tactic will fail
if unification fails. In \PE, the unification constraint can
remain partially solved.
\end{enumerate}

All these behaviours can be simulated in \PE, which is therefore a
useful framework for the study of proof-search strategies in type 
theory and for comparison with the work of 
Jojgov \cite{jojgov}, 
McBride \cite{mcbride} 
 and 
Delahaye \cite{delahaye}. 


\section{Example: commutativity of conjunction}
\label{sec:example}

We now give an example of proof-search (first introduced
in~\cite{LDMcK06} without using meta-variables)
in the \PTSC{} equivalent to System $F$, i.e. the one 
given by the sets:
\begin{center} $\SCons=\{\T,\K\}$, $\Sort=\{(\T,\K)\}$, and
$\Rel=\{(\T,\T),(\K,\T)\}$
\end{center}

For brevity, we omit types on $\l$-abstractions, abbreviate
$\cont x \el$ as $x$ for any variable $x$ and simplify $\cuti A N x P$
to $P$ when $x\not\in\FV P$.  We also write $A\wedge B$ 
   in place of its System F representation as 
\mbox{$\P Q \T {(A\arr (B\arr Q))\arr Q}$}. 

Proof-search in system \PSS\ would result in the following derivation:
$$
\Infer[\Pir]{\DeriPS {A:\T,B:\T} {\lami x{}{\lami Q{}{\lami y{} {\cont y {\st{N_B}{\st{N_A}\el}}}}}} {(A\wedge
B)\arr(B\wedge A)}}
     {
        \infer[\contr y]{\DeriPS {\Gamma} {\cont y {\st{N_B}{\st{N_A}\el}}} {Q}}
        {
                \infer  [\Pil]{\DeriPSs {\Gamma} {B\arr (A\arr Q)} {\st{N_B}{\st{N_A}\el}} {Q}}
                {
                \infer{\DeriPS \Gamma {N_B} B}
                        {\pi_B}
                \quad
                \infer[\Pil]{ \DeriPSs {\Gamma} {A\arr Q} {\st{N_A}\el} {Q}}
                        {
                        \infer{\DeriPS \Gamma {N_A} A}
                                {\pi_A}
                        \quad\infer[\ax]{ \DeriPSs {\Gamma} {Q} {\el} {Q}}{}}
                }
        }
}
$$
\noindent where $\Gamma=A:\T,B:\T,x:A\wedge B,Q:\T,y:B\arr (A\arr Q)$, 
\noindent and $\pi_A$
is the following derivation ($N_A={\cont x {\st A{\st{(\lami{x'}{}{\lami{y'}{}{x'}})}\el}}}$):
$$
\infer[\contr x]{\DeriPS \Gamma  {\cont x {\st A{\st{(\lami{x'}{}{\lami{y'}{}{x'}})}\el}}} A}
        {
        \infer[\Pil]{\DeriPSs \Gamma {A\wedge B} {\st A{\st{(\lami{x'}{}{\lami{y'}{}{x'}})}\el}} A}
                {
                \infer[\contr A]{\DeriPS \Gamma A \T}
                        {
                        \infer[\ax]{\DeriPSs \Gamma \T \el \T}{}
                        }
                \infer[\Pil]{\DeriPSs \Gamma {(A\arr (B\arr A))\arr A}{\st{(\lami{x'}{}{\lami{y'}{}{x'}})}\el} A}
                        {
                        \Infer[\Pir]{\DeriPS \Gamma {\lami{x'}{}{\lami{y'}{}{x'}}}{A\arr (B\arr A)}}
                                {
                                \infer[\contr {x'}]{\DeriPS {\Gamma,x'\colon A,y'\colon B}{x'}A}
                                        {\infer[\ax]{\DeriPSs
{\Gamma,x'\colon A,y'\colon B} A \el A}{}}
                                }
                        \infer[\ax]{\DeriPSs \Gamma A \el A}{}
                        }
                }
        }
$$ Similarly, $\pi_B$ has a derivation ($N_B={\cont x {\st
            B{\st{(\lami{x'}{}{\lami{y'}{}{y'}})}\el}}}$) with an
        analogous conclusion ${\DeriPS \Gamma {\cont x {\st
              B{\st{(\lami{x'}{}{\lami{y'}{}{y'}})}\el}}} B}$.

We now reconsider the above example in the light of system \PE. It
illustrates the need for delaying the search for a proof of the first premiss
of rule $\Pil$. Let\\
$$\begin{array}{ll}
\Gamma&=A:\T,B:\T,x:A\wedge B,Q:\T,y:B\arr A\arr Q\\
\alpha_A(\Gamma)&=\alpha_A(A,B,x,Q,y)\\
\alpha_B(\Gamma)&=\alpha_B(A,B,x,Q,y)\\
M'&=\lami x{}{\lami Q{}{\lami y{} {\cont y
      {\st{\alpha_B(\Gamma)}{\st{\alpha_A(\Gamma)}\el}}}}}\\
\Sigma&=(\Deri{\Gamma}{\alpha_B}{B}),(\Deri{\Gamma}{\alpha_A}{A}),(\constr
\Gamma Q Q)
\end{array}$$

We get the \PE-derivation below:
{\small
$$
\infer{\goals {(\Deri{A:\T,B:\T}\alpha{(A\wedge
B)\arr(B\wedge A)})} {(\alpha\mapsto \sigma_\Sigma(M'))}}
{\Infer{\enum {}{A:\T,B:\T} {(A\wedge
B)\arr(B\wedge A)}{M'} 
{\Sigma}}
     {
        \infer{\enum{} {\Gamma} Q {\cont y {\st{\alpha_B(\Gamma)}{\st{\alpha_A(\Gamma)}\el}}} {\Sigma}}
        {
          \infer  {\enums {}{\Gamma} {B\arr A\arr Q} Q{\st{\alpha_B(\Gamma)}{\st{\alpha_A(\Gamma)}\el}}{\Sigma}}
          {              
            \infer{\enum{} \Gamma B {\alpha_B(\Gamma)}{(\Deri{\Gamma}{\alpha_B}{B})}}
            \quad
            \infer { \enums{} {\Gamma} {{A\arr Q}} Q {\st{\alpha_A(\Gamma)}\el} {(\Deri{\Gamma}{\alpha_A}{A}),(\constr \Gamma Q Q)}}
            {
              {
                \infer{\enum{} \Gamma A {\alpha_A(\Gamma)} {(\Deri
                    \Gamma {\alpha_A} A)}}
                {}
                \quad
                {\infer{ \enums {}{\Gamma} {Q} Q {\el}{(\constr\Gamma Q
                      Q)}}{}}
                }
              }
            }
          }
}
\hspace{-3cm}
\infer{\goals {\Sigma}{\sigma_\Sigma}}
{\ldots}
}
$$
}

\noindent where $\sigma_\Sigma=(\alpha_B\mapsto \dom\Gamma.N_B,\alpha_A\mapsto
\dom\Gamma.N_A)$ is the solution to be obtained from the right premiss.

In the above derivation, we have systematically abandoned the sub-goals
and recorded them for later. The only choice we made was that of the
head-variable $y$, because it led to the production of the (solved)
unification constraint $(\constr \Gamma Q Q)$.

We now continue the proof-search with the right premiss, solving the
two sub-goals $(\Deri{\Gamma}{\alpha_B}{B})$ and $(\Deri{\Gamma}{\alpha_A}{A})$
that have been delayed. For instance, we can now decide to solve 
$(\Deri{\Gamma}{\alpha_A}{A})$, which will eventually produce 
   the binding 
$\alpha_A\mapsto\dom\Gamma.N_A$ with $N_A={\cont x {\st A{\st{(\lami{x'y'}{}{x'})}\el}}}$, 
as follows:
\newcommand{\thesubderivation}{
\infer{\goals {(\Deri{\Gamma}{\alpha_B}{B}),\Sigma_1,\Sigma'_1,\Sigma''_1,(\constr
\Gamma Q Q)}{\sigma}}{\ldots}
}
\newcommand{\thesubderivationname}{D}
{
$$
\infer{\goals {\Sigma}
{(\alpha_B\mapsto \dom\Gamma.N_B,
  \alpha_A\mapsto \dom\Gamma.
                  {\cont x 
                      {\st{A}
                            {\st
                               {(\lami{x'y'}{}{x'})}
                               \el
                            }
                      }
                  }
)}}
{\infer
       {\enum{} \Gamma A {\cont x {\st
             {\alpha_1(\Gamma)}{\st{(\lami{x'y'}{}{\alpha'_1(\Gamma')})}\el}}} {\Sigma_1,\Sigma'_1,\Sigma''_1}}
        {
        \infer
        {\enums {} \Gamma {A\wedge B} A {\st {\alpha_1(\Gamma)}{\st{(\lami{x'y'}{}{\alpha'_1(\Gamma')})}\el}} {\Sigma_1,\Sigma'_1,\Sigma''_1}}
                {
                \infer
                      {\enum {} \Gamma \T {\alpha_1(\Gamma)}
                           {\Sigma_1}}{}
                \infer
                {\enums{} \Gamma {(A\arr B\arr \alpha_1(\Gamma))\arr \alpha_1(\Gamma)}
                            A{\st{(\lami{x'y'}{}{\alpha'_1(\Gamma')})}\el} {\Sigma'_1,\Sigma''_1}}
                        {
                            {
                        \Infer
                        {\enum{} \Gamma {A\arr B\arr \alpha_1(\Gamma)}
                                 {\lami{x'y'}{}{\alpha'_1(\Gamma')}}
                      {\Sigma'_1}}
                                {\infer{\enum{} {\Gamma'} {\alpha_1(\Gamma)}
                                 {\alpha'_1(\Gamma')}
                                 {\Sigma'_1}}
                                       {}
                                }
                          \infer
                          {\enums {} \Gamma {\alpha_1(\Gamma)} {A} \el
                              {\Sigma''_1}}
                        {}
                      }
                      }
                }
        }
{\hspace{-2.5cm}
\thesubderivationname}
}
$$
}

\noindent where
$$\begin{array}{ll}
\alpha_1(\Gamma)&=\alpha_1(A,B,x,Q,y)\\
\Sigma_1&=(\Deri \Gamma {\alpha_1} \T)\\
\Gamma'&=\Gamma,x'\col A,y'\col B\\
\alpha'_1(\Gamma)&=\alpha'_1(A,B,x,Q,y,x',y')\\
\Sigma'_1&=(\Deri{\Gamma'}{\alpha'_1}{\alpha_1(\Gamma)})\\
\Sigma''_1&=(\constr\Gamma {\alpha_1(\Gamma)}A)\\
\sigma&=(\alpha_B\mapsto \dom\Gamma.N_B,\quad\alpha_1\mapsto\dom\Gamma.
A,\quad\alpha'_1\mapsto \dom{\Gamma'}.x')
\end{array}
$$ 
and \(\thesubderivationname\) is a sub-derivation whose conclusion is as follows: 
\[
   \thesubderivation
\]

In the above derivation, we have also abandoned the generated
sub-goals. Again we made one committing choice: that of the
head-variable $x$, which led to the unification constraint
$\constr\Gamma {\alpha_1(\Gamma)}A$.  Any other choice
of head-variable would have led to a unification constraint with no
solution. Here, this fact (and the subsequent choice of $x$) can be
mechanically noticed by a simple syntactic check.

We now continue the proof-search with the right premiss. We can
decide to solve 
$(\Deri{\Gamma}{\alpha_B}{B})$, $(\Deri \Gamma {\alpha_1} \T)$, 
or $(\Deri{\Gamma'}{\alpha'_1}{\alpha_1(\Gamma)})$.
The order in which we solve $(\Deri{\Gamma}{\alpha_B}{B})$ has little
importance (the structure is similar to that of the derivation above),
but clearly we cannot solve
$(\Deri{\Gamma'}{\alpha'_1}{\alpha_1(\Gamma)})$ before we know
$\alpha_1(\Gamma)$.
Hence, we need to solve $(\Deri \Gamma {\alpha_1} \T)$
first, which will produce 
$\alpha_1\mapsto \dom\Gamma.A$:

{
$$
\infer{\goals
  {(\Deri{\Gamma}{\alpha_B}{B}),(\Deri \Gamma {\alpha_1} \T),
(\Deri{\Gamma'}{\alpha'_1}{\alpha_1(\Gamma)}),
                   (\constr \Gamma {\alpha_1(\Gamma)} A),(\constr \Gamma Q Q)}
  {\sigma}
}
{\infer{\enum{}\Gamma \T {\cont A\el} {\constr \Gamma \T \T}}
{\infer{\enums{}\Gamma\T \T {\el} {\constr \Gamma \T \T}}
{}}
\infer{
\goals
  {(\Deri{\Gamma}{\alpha_B}{B}),(\constr \Gamma \T\T),
(\Deri{\Gamma'}{\alpha'_1}{A}),
                   (\constr \Gamma {A} A),(\constr \Gamma Q Q)}
  {\sigma'}
}
{\ldots}
}
$$}

\noindent where 
$\sigma'=(\alpha_B\mapsto \dom\Gamma.N_B,\quad\alpha'_1\mapsto \dom{\Gamma'}.x')$.

In this derivation we had to inhabit $\T$. This is a fundamental step
of the proof, even when expressed with ground terms (in system \PSS)
as above. Here, having delayed the solution of
sub-goals, we are now able to infer the correct inhabitation, directly
from the unification constraint $(\constr \Gamma {\alpha_1(\Gamma)}
A)$ which we have generated previously.  Our delaying mechanism thus
avoids many situations in which the correct choice for inhabiting a
type has to be guessed in advance, anticipating the implicit constraints
that such a choice will have to satisfy at some point.  This is hardly
mechanisable and thus leads to
numerous backtrackings.

Finally we proceed to the right premiss by solving 
$(\Deri{\Gamma'}{\alpha'_1}{A})$:
{\small
$$
\infer{
\goals
  {(\Deri{\Gamma}{\alpha_B}{B}),(\constr \Gamma \T\T),
(\Deri{\Gamma'}{\alpha'_1}{A}),
                   (\constr \Gamma {A} A),(\constr \Gamma Q Q)}
  {\sigma'}
}
{
\infer{\enum{}{\Gamma'}{A}{\cont {x'}\el} {\constr
{\Gamma'} A A}}
{\infer{\enums{}{\Gamma'}A A {\el} {\constr {\Gamma'}
A A}}
{}}
\infer{
\goals
  {(\Deri{\Gamma}{\alpha_B}{B}),(\constr \Gamma \T\T),
(\constr {\Gamma'} A A),
                   (\constr \Gamma {A} A),(\constr \Gamma Q Q)}
  {(\alpha_B(\Gamma)\mapsto N_B)}
}
{\ldots}
}
$$}

In this derivation we had to inhabit $A$. Again we made one committing
choice: that of the head-variable $x'$, which led to the unification
constraint $\constr{\Gamma'} A A$.  Again, any other choice of
head-variable would have led to obvious failure, a fact which can be
mechanically noticed by a simple syntactic check.

We can then proceed with $(\Deri{\Gamma}{\alpha_B}{B})$, in a way very
similar to that for $(\Deri{\Gamma}{\alpha_A}{A})$. We get
eventually $N_B={\cont x {\st
B{\st{(\lami{x'y'}{}{y'})}\el}}}$.

Putting it all together, 
   we have used system \PE\ to produce the following proof of the 
commutativity of conjunction:
$$\Deri {A:\T,B:\T} {\lami {xQy}{} {\cont y {\st{(\cont x {\st
B{\st{(\lami{x'y'}{}{y'})}\el}})}{\st{(\cont x {\st
A{\st{(\lami{x'y'}{}{x'})}\el}})}\el}}}} {(A\wedge B)\arr(B\wedge
A)}$$ 
The system has mechanically inferred the relevant choices of the
head-variables structuring the proof-term, by finite checks and using
the unification constraints generated by delaying the solution of
sub-goals.


\section*{Conclusion and Further Work}

In this paper we have developed a framework that serves as a
good theoretical basis for proof-search in type theory.  

Proof-search tactics in natural deduction depart from the simple
bottom-up application of the typing rules; thus their readability
and usage become more complex, as illustrated in proof-assistants such
as \Coq.
Just as in propositional logic~\cite{MR1720570}, permutation-free
sequent calculi can be a useful theoretical approach to study and
design such tactics, in the hope of improving semi-automated
reasoning.

Following these ideas, we have defined a parameterised formalism
giving a sequent calculus for each \PTS. It comprises a syntax, a
rewrite system and typing rules.  In contrast to previous work, the
syntax of both types and proof-terms of $\PTSCa$ is in
sequent calculus style, thus avoiding implicit or explicit conversions
to natural deduction~\cite{gutierrez03elimination,pinto00sequent}. 
We have given a direct proof, by simulation, of confluence for each $\PTSCa$. 

   We have established a strong correspondence with natural deduction
(regarding both logic and strong normalisation), 
   when restricted to the ground terms $\PTSC$ of a given $\PTSCa$. 
These results and their proofs were 
formalised in \Coq~\cite{SilesPTSC}.  We can give as examples the
corners of Barendregt's \(\lambda\)-cube, for which we now have an elegant
theoretical framework for proof-search: We have shown how to deal with
conversion rules so that basic proof-search tactics are simply the
root-first application of the typing rules.

These ideas have then been extended, in the calculi \PTSCa, by the use of
meta-variables to formalise the notion of incomplete proofs, and their
theory has been studied.  The approach differs from~\cite{Munozsynthesis}
both in that we use sequent calculus rules, which match proof-search
tactics, and in that our system simulates $\beta$-reduction.

We have shown that, in particular, the explicit use of meta-variables
avoids the phenomenon of \emph{r-splitting} and allows for more
flexibility in proof-search, where sub-goals can be tackled in the
order that is most suitable for each situation. Such a flexibility
avoids some of the need for ``guess-work'' in proof-search, and
formalises some mechanisms of proof-search tactics in proof
assistants. This approach has been illustrated by the example of
commutativity of conjunction.

Our system does not commit to specific search strategies \emph{a
priori}, so that it can be used as a general framework to investigate
such strategies, as discussed at the end of
Section~\ref{sec:enum}. This could reflect various degrees of user
interaction in proof-search.

Ongoing work includes the incorporation of some of these ideas into
the redesign of the \Coq\ proof engine~\cite{Coq}. It also includes
the treatment of $\eta$-conversion, a feature that is currently
lacking in the \PTS-based system \Coq. We expect that, by adding
$\eta$-\emph{expansion} to our system, our approach to proof-search
can be related to that of \emph{uniform proofs} in logic programming.

Further work includes studying direct proofs of strong normalisation
(such as Kikuchi's for propositional logic~\cite{Kikuchi04}), and
dealing with inductive types such as those used in \Coq. Their
specific proof-search tactics should also clearly appear in sequent
calculus. Finally, given the importance of sequent calculi for
classical logic, it would be interesting to build classical Pure Type
Sequent Calculi.

\paragraph{Acknowledgements}
The authors are grateful to Delia Kesner, Gilles Dowek, Hugo Herbelin,
Arnaud Spiwack, Vincent Siles, Alex Simpson and David Pym for their
helpful remarks and comments, and for pointing out important items of
related work.


\bibliographystyle{Common/good-url}
\bibliography{Common/abbrev,Common/Main,Common/crossrefs}

\appendix


\theoremstyle{plain}

\newtheorem{athm}{Theorem}
\newtheorem{acor}[athm]{Corollary}
\newtheorem{alem}[athm]{Lemma}
\newtheorem{aprop}[athm]{Proposition}
\newtheorem{aasm}[athm]{Assumption}

\theoremstyle{definition}

\newtheorem{arem}[athm]{Remark}
\newtheorem{arems}[athm]{Remarks}
\newtheorem{aexa}[athm]{Example}
\newtheorem{aexas}[athm]{Examples}
\newtheorem{adefi}[athm]{Definition}
\newtheorem{aconv}[athm]{Convention}
\newtheorem{aconj}[athm]{Conjecture}
\newtheorem{aprob}[athm]{Problem}
\newtheorem{aoprob}[athm]{Open Problem}
\newtheorem{aalgo}[athm]{Algorithm}
\newtheorem{aobs}[athm]{Observation}
\newtheorem{afact}[athm]{Fact}
\newtheorem{aqu}[athm]{Question}
\newtheorem{apty}[athm]{Property}
\newtheorem{aoqu}[athm]{Open Question}

\section*{Subject Reduction}

\begin{adefi}
We write $\Derist \Gamma M A$ (resp. $\Derisst \Gamma B l C$)
whenever we can derive $\Deri \Gamma M A$ (resp. $\Deris \Gamma B
l C$) and the last rule is not a conversion rule.
\end{adefi}

The following Lemma is easily derived by induction on the typing tree:
\begin{alem}[Generation Lemma]$ $\label{Lem:Generation}
\begin{enumerate}[\em(1)]
\item
\begin{enumerate}[\em(a)]
\item
If $\DeriPTSC\Gamma {s} C$ then there is $s'$ such that $\Derist\Gamma {s} {s'}$
with $C\conv s'$.
\item
If $\DeriPTSC\Gamma {\P x A B} C$ then there is $s$ such that $\Derist\Gamma {\P x A B} s$
with $C\conv s$.
\item If
$\DeriPTSC\Gamma {\lami x A M} C$ then\\ there is $B$ such that
$C\conv{\P x A B}$ and $\Derist\Gamma {\lami x A M}{\P x A B}$.
\item
If $\DeriPTSC\Gamma {\cuti A M x N} C$ then there is $C'$ such that $\Derist\Gamma {\cuti A M
x N} {C'}$ with $C\conv {C'}$.
\item
If $M$ is not of the above forms and $\DeriPTSC\Gamma M C$, then $\Derist\Gamma M C$.
\end{enumerate}
\item
\begin{enumerate}[\em(a)]
\item
If $\DeriPTSCs\Gamma B {\el} C$ then $B\conv C$.
\item
If $\DeriPTSCs\Gamma D {\st M l} C$ then\\ there are $A,B$ such that $
D\conv {\P x A B}$ and $\Derisst\Gamma {\P x A B} {\st M l} C$.
\item
If $\DeriPTSCs\Gamma B{\cuti A M x l} C$ then are $B', C'$ such that\\
$\Derisst\Gamma
{B'}{\cuti A M x l} {C'}$ with $ C\conv{C'}$ and $ B\conv{B'}$.
\item
If $l$ is not of the above forms and $\DeriPTSCs\Gamma D {l} C$ then
$\Derisst\Gamma D {l} C$.
\end{enumerate}
\end{enumerate}
\end{alem}
   \proof 
Straightforward induction on the typing tree.
   \qed

\begin{arem}
The following rule is derivable, using a conversion rule:
$$
\Infers{\DeriPTSC \Gamma Q A\qquad\DeriPTSC {\Gamma,(x:A),\Delta} {M}
C
\quad \DeriPTSC{\Delta'}{\cuti A Q x {C}}s\quad \Gamma,\cuti A Q x {\Delta}\subenv {\Delta'}}
{\DeriPTSC {{\Delta'}}
{\cuti A Q x {M}} {\cuti A Q x {C}}}
{} 
$$
\end{arem}

Proving subject reduction relies on the following properties of $\Rew{\B\x}$:
\begin{alem}\hfill
\begin{enumerate}[$\bullet$]
\item
Two distinct sorts are not convertible.
\item
A $\Pi$-construct is not convertible to a sort.
\item
$\P x A B\conv\P x D E$ if and only if $A\conv D$ and $B\conv
E$.
\item
If $y\not\in FV(P)$, then $P\conv\cuti A N y P$.
\item
$\cuti K M y{\cuti K N x P}\conv \cuti K {\cuti K M y N} x {\cuti
K M y P}$ (provided $x\not\in FV(M)$).
\end{enumerate}
\end{alem}
   \proof
The first three properties are a consequence of the confluence of the
rewrite system (Corollary~\ref{cor:confluencePTSC}). The last two rely on
the fact that the system \xsubst {} is terminating, so that only the
case when $P$ is an \xsubst-normal form remains to be checked, which
is done by structural induction.
   \qed 

Using all of the results above, subject reduction can be
proved:

\begin{athm}[Subject reduction in a \PTSC]$ $\label{th:SR-app}
\begin{enumerate}[\em(1)]
\item
If $\DeriPTSC \Gamma M X$ and $M\Rew{B\x} M'$, 
then $\DeriPTSC \Gamma {M'} X$
\item
If $\DeriPTSCs \Gamma Y l Z$ and $l\Rew{B\x} l'$, 
then $\DeriPTSCs \Gamma Y {l'} Z$
\end{enumerate}
\end{athm}

   \proof
By simultaneous induction on the typing tree. For every rule, if the
reduction takes place within a sub-term that is typed by one of the
premisses of the rule (\eg the conversion rules), then we can apply
the induction hypothesis on that premiss. In particular, this takes
care of the cases where the last typing rule is a conversion rule.

So it now suffices to look at the root reductions. For lack of space
we often do not display some minor premisses in following
derivations, but we mention them before or after. We also drop the
subscript \PTSC\ from derivable statements.
\begin{enumerate}[\textsf{As}]
\item[\textsf{B}]
$\cont{(\lami x A N)}{(\st P {l_1})} \Rew{} \cont{(\cuti A P x N)}{{l_1}}$

\noindent By the Generation Lemma, 1.(c) and 2.(b), there exist $B$, $D$, $E$
such that:
{\normalsize
$$
\infers {
    \infers{\Deri \Gamma {\P x A B} {s}\quad\Deri{\Gamma,x:A}{N}{B}}
            {\Deri\Gamma{\lami x A N}{C}}
            {}
    \quad
    \infers{\Deri \Gamma P D\quad \Deris\Gamma{\cuti D P x E}{l_1} X}
            {\Deris\Gamma C{\st P {l_1}}X}
            {}
    }
    {\Derist\Gamma{\cont{(\lami x A N)}{(\st P {l_1})}}{X}}
    {}
$$
}with ${\P x A B}\conv  C \conv \P x D E$. Therefore, $A\conv D$ and
$B\conv E$.
Moreover, $\Deri \Gamma {A} {s_A}$, $\Deri {\Gamma,x:A} {B}
{s_B}$ and $\Gamma\wf $.
Hence, we obtain 
$\Deri \Gamma {\cuti A P x B}{s_B}$, so:
{\normalsize
$$
\infers{
\infers{
    \infers{\Deri \Gamma P D}
        {\Deri \Gamma P A}{}
    \quad
    \Deri{\Gamma,x:A}{N}{B}
        }
{\Deri \Gamma {\cuti A P x N}{\cuti A P x B}}
{}
\quad \infers{\Deris\Gamma{\cuti D P x E}{l_1} X}
{\Deris\Gamma{\cuti D P x B}{l_1} X}{}}
{\Deri{\Gamma}{(\cont{\cuti A P x N}{l_1})}{X}}
{}
$$
}with $\cuti A P x B\conv \cuti A P x E$.
\item[\As]
\begin{enumerate}[\textsf{A1}]
\item[\textsf{A1}]
$\conc{(\st N {l_1})} {l_2}\Rew{} \st N {(\conc {l_1} {l_2})}$

By the Generation Lemma 2.(b), there are $A$ and $B$ such that $Y\conv \P x A B$ and:
{\normalsize
$$
\infers{
    \infers{
        \Deri \Gamma {\P x A B} s\quad
        \Deri \Gamma N A\quad
        \Deris \Gamma {\cuti A N x B} {l_1} C
        }
        {\Deris \Gamma {Y} {\st N {l_1}} C}{} \quad
        \Deris{\Gamma}{C}{l_2}{Z}
    }
    {\Derisst{\Gamma}{Y}{\conc {(\st N {l_1})} {l_2}}{Z}}
    {}
$$
}Hence,
{\normalsize
$$
\infers{\Deri \Gamma Y{s_Y}\;\infers{
        \Deri \Gamma {\P x A B} s\quad
        \Deri \Gamma N A\quad
        \infers{
            \Deris \Gamma {\cuti A N x B} {l_1} C
            \quad
            \Deris{\Gamma}{C}{l_2}{Z}
            }
        {\Deris \Gamma {\cuti A N x B} {\conc {l_1} {l_2}} Z}
        {}
        }
    {\Deris{\Gamma}{\P x A B}{\st N {(\conc {l_1} {l_2})}}{Z}}{}
}
{\Deris{\Gamma}{Y}{\st N {(\conc {l_1} {l_2})}}{Z}}{}
$$
}
\item[\textsf{A2}]
$\conc{\el} {l_1}\Rew{} l_1$

By the Generation Lemma 2.(a), we have $A\conv Y$ and
{\normalsize
$$
\infers{\Deris \Gamma Y \el A\quad\Deris{\Gamma}{A}{l_1}{Z}}{\Derisst{\Gamma}Y{\conc
{\el} {l_1}}{Z}}{}
$$
}
Since $\Deri \Gamma Y{s_Y}$, we obtain
{\normalsize
$$
\infers{
\Deris \Gamma A {l_1} Z
    }
    {\Deris \Gamma Y {l_1} Z}
    {}
$$
}
\item[\textsf{A3}]
$\conc{(\conc{l_1}{l_2})} {l_3}\Rew{} \conc{l_1} {(\conc{l_2}{l_3})}$

By the Generation Lemma 2.(d),
{\normalsize
$$
\infers{\infers{\Deris\Gamma Y{l_1}B
\quad\Deris\Gamma B{l_2}A}{\Derisst\Gamma Y{\conc{l_1}{l_2}}A}{}\quad\Deris{\Gamma}{A}{l_3}{Z}}
{\Derisst{\Gamma}Y{\conc{(\conc{l_1}{l_2})} {l_3}}{Z}}{}
$$
}Hence,

$$
\infers{\Deris\Gamma Y{l_1}B
\quad\infers{\Deris\Gamma B{l_2}A
\quad \Deris{\Gamma}{A}{l_3}{Z}}{\Deris\Gamma B{\conc{l_2}{l_3}}Z}{}}
{\Deris{\Gamma}Y{\conc{l_1} {(\conc{l_2}{l_3})}}{Z}}{}
$$
\end{enumerate}
\item[\Bs]
\begin{enumerate}[\textsf{B1}]
\item[\textsf{B1}]
$\cont N \el\Rew{} N$\\
{\normalsize
$$
\infers{\Deri \Gamma N A\quad\Deris{\Gamma}{A}{\el}{X}}{\Derist{\Gamma}{\cont
{N} {\el}}{X}}{}
$$
}By the Generation Lemma 2.(a), we have $A\conv X$.\\
Since $\Deri \Gamma X{s_X}$, we obtain
{\normalsize
$$
\infers{
\Deri \Gamma N A
    }
    {\Deri \Gamma N X}
    {}
$$
}
\item[\textsf{B2}]
$\cont {(\cont x {l_1})} {l_2}\Rew{} \cont x {(\conc {l_1} {l'})}$

By the Generation Lemma 1.(e),
{\normalsize
$$
\infers{\infers{ \Deris{\Gamma}{A}{l_1}{B}\quad
(x:A)\in\Gamma}{\Derist \Gamma {\cont x
l}{B}}{}\quad\Deris{\Gamma}{B}{l_2}{X}}{\Derist{\Gamma}{\cont
{(\cont x {l_1})} {l_2}}{X}}{}
$$
}Hence,
{\normalsize
$$
\infers{
    (x:A)\in\Gamma\quad
    \infers{
        \Deris{\Gamma}{A}{l_1}{B}\quad
        \Deris{\Gamma}{B}{l_2}{X}
        }
        {\Deris \Gamma A {\conc {l_1}{l_2}}{X}}
        {}
    }
    {\Deri{\Gamma}{\cont x {(\conc {l_1}{l_2})}}{X}}
    {}
$$
}
\item[\textsf{B3}]
$\cont {(\cont N {l_1})} {l_2}\Rew{} \cont N {(\conc {l_1} {l_2})}$

By the Generation Lemma 1.(e),
{\normalsize
$$
\infers{
    \infers{
        \Deri {\Gamma} N A\quad
        \Deris{\Gamma}{A}{l_1}{B}
        }
        {\Derist \Gamma {\cont N {l_1}}{B}}
        {}\quad
        \Deris{\Gamma}{B}{l_2}{X}
    }
    {\Derist{\Gamma}{\cont {(\cont N {l_1})} {l_2}}{X}}
    {}
$$
}Hence,
{\normalsize
$$
\infers{
    \Deri {\Gamma} N A\quad
    \infers{
        \Deris{\Gamma}{A}{l_1}{B}\quad
        \Deris{\Gamma}{B}{l_2}{X}
        }
        {\Deris \Gamma A {\conc {l_1}{l_2}}{X}}
        {}
    }
    {\Deri{\Gamma}{\cont N {(\conc {l_1}{l_2})}}{X}}
    {}
$$
}
\end{enumerate}
\item[\Cs]
We have a redex of the form $\cuti A Q y {R}$ typed by:
{\normalsize
$$
\infers{
        {\Deri{\Delta'}{Q}E}{}
    \quad
        {\Deri{{\Delta'},y:E,\Delta}{R}{X'}}{}
    \quad
        {\Delta'},\cuti G {Q} y \Delta \subenv \Gamma\wf 
        }
{\Derist\Gamma{\cuti G {Q} y {R}}X}{}
$$
}with either $X=X'\in\SCons$ or $X=\cuti A Q y {X'}$. \\
In the latter case,
$\Deri{\Gamma} X {s_X}$ for some $s_X\in\SCons$. We also have
${\Gamma}\wf $.

Let us consider each rule:
\begin{enumerate}[\textsf{C1}]
\item[\textsf{C1}]
$\cuti G Q y {\lami x A N}\Rew{} \lami x {\cuti G Q y  A} {\cuti G Q y
  N}$

$R={\lami x A N}$\\
By the Generation Lemma 1.(b), there is $s_3$ such that $C\conv s_3$ and:
{\normalsize
$$
\infers{
    \infers{\Deri {{\Delta'},y:E,\Delta} A {s_1}\qquad \Deri {{\Delta'},y:E,\Delta,x:A} B {s_2}}
    {\Deri {{\Delta'},y:E,\Delta} {\P x A B} {C}}{}
    \Deri
{{\Delta'},y:E,\Delta,x:A} N {B}}
    {\Deri
{{\Delta'},y:E,\Delta} {\lami x A N} {X'}}{}
$$
}with $(s_1,s_2,s_3)\in\Rel$ and $X'\equiv \P x A B$.
Therefore,
$X'\not\in\SCons$, and as a consequence $X=\cuti E Q y X'\conv\cuti E Q y {\P x {A}
B}\conv \P x {\cuti E Q y A} {\cuti E Q y B}$. We have: 
{\normalsize
$$
\infers{\Deri{\Delta'}{Q}E\quad\Deri {{\Delta'},y:E,\Delta} A {s_1}}{\Deri {\Gamma} {\cuti E Q y A} {s_1}}{}
$$
}Hence, $\Gamma,x:\cuti E Q y A\wf $ and ${\Delta'},\cuti G {Q} y
\Delta,x:\cuti G {Q} y A \subenv \Gamma,x:\cuti E Q y A$, so:
{\normalsize
$$
\infers{\Deri{\Delta'}{Q}E\quad\Deri {{\Delta'},y:E,\Delta,x:A} B {s_2}}
{\Deri {\Gamma,x:\cuti E Q y A} {\cuti E Q y B} {s_2}}{}
$$
}so that $\Deri \Gamma {\P x {\cuti E Q y A} {\cuti E Q y B}}
{s_3}$ and
{\normalsize
$$
\infers{
\infers{
    \Infer
    {\Deri{\Gamma,x:\cuti E Q y A}{\cuti E Q y N}{\cuti E Q y B}
        }{\Deri{\Delta'}{Q}E\quad\Deri
{{\Delta'},y:E,\Delta,x:A} N {B}}{}}
{\Deri{\Gamma}{\lami x{\cuti E Q y A}{\cuti E Q y N}}{\P x
{\cuti E Q y A} {\cuti E Q y B}}}{}
\quad X \conv {\P x
{\cuti E Q y A} {\cuti E Q y B}}
}
{\Deri{\Gamma}{\lami x{\cuti E Q y A}{\cuti E Q y N}} X}
{}
$$
}\item[\textsf{C2}]
$\cuti G Q y {(\cont y {l_1})}\Rew{} \cont Q {\cuti G Q y  {l_1}}$

$R={\cont y {l_1}}$\\
By the Generation Lemma 1.(e), $\Deris {{\Delta'},y:E,\Delta} E {l_1} {X'}$. Now notice that $y\not\in
FV(E)$, so $\cuti E Q y E\conv E$ and $\Deri {\Delta'} E {s_E}$. Also,
${\Delta'} \subenv\Gamma$, so
{\normalsize
$$
\infers{
    \winfer{\Deri {\Gamma} Q E}{\Deri {{\Delta'}} Q E}
    {}\quad \infers{
\infers{\Deri {\Delta'} Q E\quad\Deris {{\Delta'},y:E,\Delta} E {l_1} {X'}}{\Deris {\Gamma} {\cuti G Q y E} {\cuti
G Q y {l_1}} {X}}
 {}\quad \winfer{\Deri \Gamma E {s_E}}{\Deri {\Delta'} E {s_E}}{}}
 {\Deris {\Gamma} {E} {\cuti
G Q y {l_1}} {X}}
 {}} {\Deri \Gamma {\cont Q {\cuti G Q y {l_1}}}{X}}{}
$$
}\item[\textsf{C3}]
$\cuti G Q y {(\cont x {l_1})}\Rew{} \cont x {\cuti G Q y  {l_1}}$

$R={\cont x {l_1}}$\\
By the Generation Lemma 1.(e), \mbox{$\Deris {{\Delta'},y:E,\Delta} A {l_1} {X'}$}
with  $(x:A)\in {\Delta'},\Delta$. Let $B$ be the type of $x$
in $\Gamma$. We have
{\normalsize
$$
\infers{
\infers{
\infers{\Deri {\Delta'} Q E\quad\Deris {{\Delta'},y:E,\Delta} A {l_1} {X'}}{\Deris {\Gamma} {\cuti G Q y A} {\cuti G Q
y {l_1}} {X}}
 {}\quad \Deri\Gamma B {s_B}}
 {\Deris {\Gamma} {B} {\cuti G Q y {l_1}}
{X}}
 {}} {\Deri {\Gamma} {\cont x
{\cuti G Q y {l_1}}} {X}}{}
$$}Indeed, if
$x\in\dom\Delta$ then $B\conv\cuti E Q y A$, otherwise $B\conv A$
with $y\not\in FV(A)$, so in each case $B\conv\cuti E Q y A$.
Besides, $\Gamma\wf $ so $\Deri\Gamma B {s_B}$.
\item[\textsf{C4}]
$\cuti G Q y {(\cont N {l_1})}\Rew{} \cont {\cuti G Q y  N} {\cuti G Q
  y  {l_1}}$

$R={\cont N {l_1}}$\\
By the Generation Lemma 1.(e),
{\normalsize
$$\infers{\Deri {{\Delta'},y:E,\Delta} N {A}\quad\Deris {{\Delta'},y:E,\Delta} A {l_1} {X'}}
{\Derist {{\Delta'},y:E,\Delta} {\cont N {l_1}} {X'}}{}
$$
}Also, we have
{\normalsize
$$
\infers{\Deri {\Delta'} Q E\quad\Deri {{\Delta'},y:E,\Delta} {A} {s_A}}{\Deri {\Gamma}
{\cuti G Q y A} {s_A}}{}
$$
}Hence,
{\normalsize
$$\infers{
    \infers{\Deri {\Delta'} Q E\quad\Deri {{\Delta'},y:E,\Delta} N {A}}{\Deri {\Gamma} {\cuti G Q y N} {\cuti G Q y A}}{}
    \quad
    \infers{\Deri {\Delta'} Q E\quad\Deris {{\Delta'},y:E,\Delta} A {l_1} {X'}}{\Deris {\Gamma} {\cuti G Q y A} {\cuti G Q y {l_1}} {X}}{}}
{\Deri {\Gamma} {\cont {\cuti G Q y N} {\cuti G Q y {l_1}}} {X}}{}$$}
\item[\Cv]
$\cuti G Q y {\P x A B}\Rew{} \P x {\cuti G Q y  A} {\cuti G Q y B}$

$R={\P x A B}$\\
By the Generation Lemma 1.(b), there exists $s_3$ such that $X'\conv s_3$ and:
{\normalsize
$$
    \infers{\Deri {{\Delta'},y:E,\Delta} A {s_1}\qquad \Deri {{\Delta'},y:E,\Delta,x:A} B {s_2}}
    {\Deri {{\Delta'},y:E,\Delta} {\P x A B} {X'}}{}
$$}
with $(s_1,s_2,s_3)\in\Rel$.
{\normalsize
$$
\infers{\Deri{\Delta'}{Q}E\quad\Deri {{\Delta'},y:E,\Delta} A {s_1}}{\Deri {\Gamma} {\cuti E Q y A} {s_1}}{}
$$
}Hence, $\Gamma,x:\cuti E Q y A\wf $ and
${\Delta'},\cuti G {Q} y
\Delta,x:\cuti G {Q} y A \subenv \Gamma,x:\cuti E Q y A$, so we obtain:
{\normalsize
$$
\infers{\Deri{\Delta'}{Q}E\quad\Deri {{\Delta'},y:E,\Delta,x:A} B {s_2}}
{\Deri {\Gamma,x:\cuti E Q y A} {\cuti E Q y B} {s_2}}{}
$$
}and hence that $\Deri \Gamma {\P x {\cuti E Q y A} {\cuti E Q y B}}
{s_3}$. 
\\Now if $X'\in\SCons$, then $X=X'=s_3$ and we
are done. 
\\Otherwise $X=\cuti K Q y {X'}\conv \cuti K Q y
{s_3}\conv s_3$, and we conclude using a conversion rule (because
$\Deri \Gamma X {s_X}$).
\item[\textsf{C6}]
$\cuti G Q y {s}\Rew{} s$
and 
$R=s$. 
By the Generation Lemma 1.(a), we obtain $X'\conv s'$ for some $s'$ with $(s,s')\in\Sort$.
Since $\Gamma\wf $, we obtain $\Deri \Gamma s {s'}$.
If $X'\in\SCons$, then $X=X'=s'$ and we are done. 
Otherwise $X=\cuti K Q y {X'}\conv \cuti K Q y
{s'}\conv s'$ and we conclude
using a conversion rule (because $\Deri \Gamma X {s_X}$).
\end{enumerate}
\item[\Ds]
We have a redex of the form $\cuti A Q y {l_1}$ typed by:
{\normalsize
$$
\infers{
        {\Deri{\Delta'}{Q}E}{}
    \quad
        {\Deris{{\Delta'},y:E,\Delta}{Y'}{l_1}{Z'}}{}
    \quad
        {\Delta'},\cuti G {Q} y \Delta \subenv \Gamma\wf 
        }
{\Derisst\Gamma{Y}{\cuti G {Q} y {l_1}}Z}{}
$$
}with \mbox{$Z=\cuti A Q y {Z'}$} and \mbox{$Y=\cuti G {Q} y {Y'}$}. 
We also have ${\Gamma}\wf $, $\Deri \Gamma Y{s_Y}$ and $\Deri \Gamma Z{s_Z}$.

Let us consider each rule:

\begin{enumerate}[\textsf{D1}]
\item[\textsf{D1}]
$\cuti G Q y \el \Rew{} \el$

$l_1={\el}$\\
By the Generation Lemma 2.(a), $Y'\conv X'$, so $Y\conv X$.
{\normalsize
$$
    \infers{
    \infers{\Deri \Gamma Y{s_Y}}{\Deris \Gamma Y \el Y}{}
    \qquad \Deri \Gamma X{s_X}}
    {\Deri {Y} {\el} {X}}{}
$$
}\item[\textsf{D2}]
$\cuti G Q y {(\st N {l_2})}\Rew{} \st {(\cuti G Q y {N})} {(\cuti
G Q
y {l_2})}$

$l_1={\st N {l_2}}$\\
By the Generation Lemma 2.(b), there are $A$, $B$ such that\linebreak $Y'\conv \P
x A B$ and:
{\normalsize
$$
\infers{\Deri {{\Delta'},y:E,\Delta} {\P x A B} s\quad \Deri
{{\Delta'},y:E,\Delta} N A\quad\Deris{{\Delta'},y:E,\Delta}{\cuti G N x
B}{l_2}{Z'}}{\Derisst{{\Delta'},y:E,\Delta}{\P x A B}{l_1}{Z'}}{}
$$}From $\Deris{{\Delta'},y:E,\Delta}{\cuti G N x
B}{l_2}{Z'}$ we obtain 
$$
\Deris{\Gam}{\cuti E Q y {\cuti G N x
B}}{\cuti E Q y {l_2}}{Z}
$$
From $\Deri {{\Delta'},y:E,\Delta} N A$ we obtain $\Deri
{\Gam} {\cuti E Q y N} {\cuti E Q y A}$.\\
From $\Deri {{\Delta'},y:E,\Delta} {\P x A B} s$ part (b) of the Generation
Lemma 1 allows us to conclude
$\Deri {{\Delta'},y:E,\Delta} {A} {s_A}$ and 
$\Deri {{\Delta'},y:E,\Delta,x\col A} {B} {s_B}$.
Hence we obtain 
$$
\infer{\Deri {\Gam} {\cuti E Q y A} {s_A}}
{\Deri {{\Delta'},y:E,\Delta} {A} {s_A}}
$$
and thus $\Gam,x\col \cuti E Q y A\wf $ and then
$$
\infer{\Deri {\Gam,x\col \cuti E Q y A} {\cuti E Q y B} {s_B}}
{\Deri {{\Delta'},y:E,\Delta,x\col A} {B} {s_B}}
$$
From that we obtain both
$\Deri {\Gamma} {\P x {\cuti E Q y A} {\cuti E Q y{B}}} s $
and\\
$\Deri {\Gam} {\cuti E {\cuti E Q y N}x{\cuti E Q y B}} {s_B}$.\\
Note that $\P x {\cuti E Q y A} {\cuti E Q y{B}}\conv{\cuti
E Q y{\P x A B}}\conv{\cuti E Q y{Y'}}=Y$.
 We
obtain 
{\normalsize
$$
\infers
{\infers{
\Deri {\Gamma} {\cuti E Q y N} {\cuti E Q y A}\quad
\infers{\Deris{\Gamma}{\cuti E Q y {\cuti G N x B}}{\cuti E Q y
{l_2}}{Z}}{\Deris{\Gamma}{\cuti G {\cuti E Q y N} x {\cuti E Q y
B}}{\cuti E Q y {l_2}}{Z}}{}}{\Deris \Gamma {\P x {\cuti E Q y A}
{\cuti E Q y{B}}} {\st {(\cuti G Q y {N})} {(\cuti G Q y {l_2})}}
Z}{}}
{\Deris \Gamma 
{Y} {\st {(\cuti G Q y {N})} {(\cuti G Q y {l_2})}}
Z}{}
$$
}\item[\textsf{D3}]
$\cuti G Q y {(\conc {l_2}{l_3})}\Rew{} \conc {(\cuti G Q y
{l_2})}
{(\cuti G Q y {l_3})}$ 

$l_1=\conc {l_2}{l_3}$\\
By the Generation Lemma 2.(d),
{\normalsize
$$
\infers{\Deris {{\Delta'},y:E,\Delta} {Y'}{l_2}
A\quad\Deris{{\Delta'},y:E,\Delta}{A}{l_3}{Z'}}{\Derisst{{\Delta'},y:E,\Delta}{Y'}{\conc
{l_2}{l_3}}{Z'}}{}
$$
}Hence,
{\normalsize
$$
    \infers{
\Deris {\Gamma} Y {\cuti E Q y {l_2}} {\cuti E Q y A}\qquad
\Deris{\Gamma}{\cuti E Q y {A}}{\cuti E Q y {l_3}}{Z}}{\Deris
\Gamma {Y} {\conc {(\cuti G Q y {l_2})} {(\cuti G Q y {l_3})}}
Z}{}\eqno{\qEd}
$$
}\end{enumerate}
\end{enumerate}
\vspace{-20 pt}


\end{document}